\documentclass[aps,preprint]{revtex4}
\usepackage{epsfig}
\usepackage{graphicx}
\usepackage{amsmath}
\usepackage{amssymb}

\begin{document}

\title{Soft lubrication: the elastohydrodynamics of non-conforming and conforming contacts}
\author{J.M. Skotheim$^{1,2}$ \& L. Mahadevan$^{1,2}$}
\affiliation{ $^1$Department of Applied Mathematics and Theoretical Physics, Centre for Mathematical Sciences, Cambridge CB3 0WA, UK.\\
$^2$Division of Engineering and Applied Sciences, Harvard University, Pierce Hall, 29 Oxford Street,
Cambridge, MA 02138, USA}

\date{\today}

\begin{abstract}
We study the lubrication of fluid-immersed soft interfaces and  show that elastic deformation couples tangential and normal forces and thus generates lift.  We consider materials that deform easily, due to either geometry ({\it e.g} a shell) or constitutive properties ({\it e.g.} a gel or a rubber), 
so that the effects of pressure and temperature on the fluid properties may be neglected.
Four different system geometries are considered:  a rigid cylinder moving parallel to a soft layer coating a rigid substrate; a soft cylinder moving parallel to a rigid substrate; a cylindrical shell moving parallel to a rigid substrate; and finally a cylindrical conforming journal bearing coated with a thin soft layer.  
In addition, for the particular case of a soft layer coating a rigid substrate we consider both elastic and poroelastic material responses.  
For all these cases we find the same generic behavior: there is an optimal combination of geometric and material parameters that maximizes the dimensionless normal force as a function of the softness parameter
$\eta = \frac{\hbox{hydrodynamic pressure}}{\hbox{elastic stiffness}}  = \frac{\hbox{surface deflection}}{\hbox{gap thickness}} $ which characterizes the fluid-induced deformation of the interface. The corresponding cases for a spherical slider are treated using scaling concepts.  
\end{abstract}

\maketitle

\section{Introduction}

The reduction of friction in practical applications has been studied since antiquity.  
Pictographs found in Uruk, located in modern day Iraq, have been dated to ca. 3000 B.C. and 
illustrate the transition from sleds to wheels (see Hamrock \& Dowson (1981) for a historical overview).
While this advance certainly reduced friction, further reductions were possible upon the introduction 
of a viscous lubricating fluid in the axle joints.  The theoretical underpinings of fluid lubrication in such geometries can be traced back to the work of Reynolds (1886), who studied the mechanics of fluid flow through a thin gap using an  approximation to the Stokes' equations, now known as lubrication theory.  Recent efforts in this technologically important problem have focused on modifications 
of Reynolds' lubrication theory to account for elastohydrodynamic effects (elastic surface deformation due to fluid pressure), piezoviscous behavior (lubricant viscosity change due to high pressure),  and thermoviscous behavior (lubricant viscosity change due to frictional heating) 
(Dowson \& Higginson 1959; O'Donoghue, Brighton \&Hooke 1967; Conway \& Lee 1975; 
Hamrock \& Dowson 1981).

Inspired by a host of applications in physical chemistry, polymer physics and biolubrication, in this paper we focus on the elastohydrodynamics of soft interfaces, which deform easily thereby precluding piezoviscous and thermoviscous effects. There have been a number of works in these areas in the context of specific problems such as cartilage biomechanics (Grodzinsky, Lipshitz \& Glimcher 1978; Mow, Holmes \& Lai 1984; Mow \& Guo 2002), the motion of red blood cells in capillaries
(Lighthill 1968; Fitz-Gerald 1969; T\"ozeren \& Skalak 1978; Secomb {\it et al.} 1986; Damiano {\it et al.} 1996;  Secomb, Hsu \& Pries 1998; Feng \& Weinbaum 2000; Weinbaum {\it et al.} 2003), 
the elastohydrodynamics of rubber-like elastomers (Tanner 1966; Martin {\it et al.} 2002), 
polymer brushes (Klein, Perahia \& Warburg 1991; Sekimoto \& Leibler 1993) 
and vesicles (Abkarian, Lartigue \& Viallat 2002; Beaucourt, Biben \& Misbah 2004). Another related phenomenon is that of a bubble rising slowly near a wall;  the
bubble's surface deformation then leads to a lift force (Leal 1980; Takemura {\it et al.} 2002). Instead of focusing on specific applications, here we address a slightly different set of questions:  How can one generate lift between soft sliding surfaces to increase separation and reduce wear?
What is the role of geometry in determining the behavior of such systems? How do material properties influence the elastohydrodynamics? Are there optimal combinations of the geometry and material properties that maximize the lift force? And finally, can the study of soft lubrication lead to improved engineering designs and be of relevance to real biological systems? 

\begin{figure}
\begin{center}
\includegraphics[width=10cm]{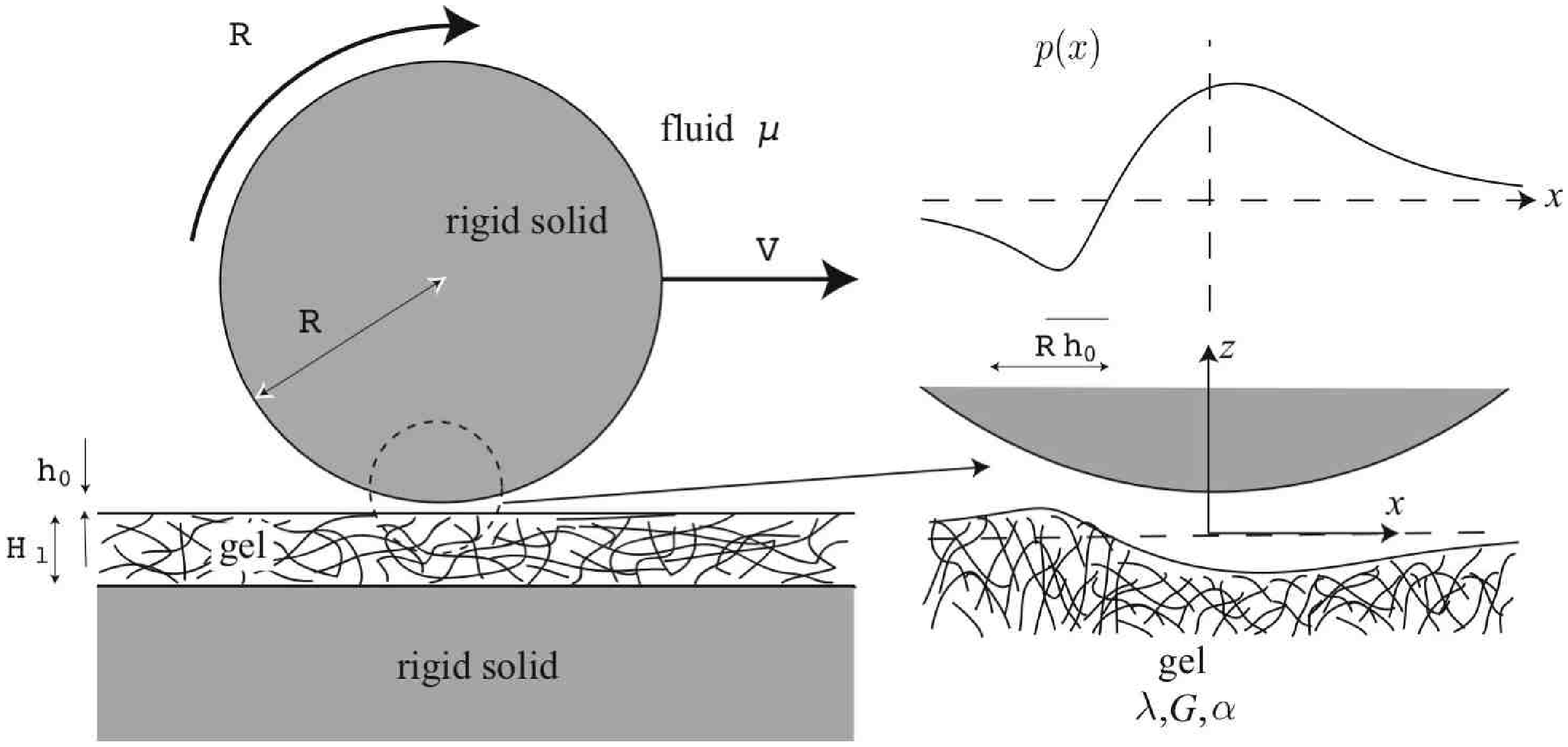}
\par
\caption{A solid cylinder moves through a liquid of viscosity $\mu$ above a thin gel layer of thickness
$H_l$ that covers 
a rigid solid substrate.  The asymmetric pressure distribution pushes down on the gel when
the fluid pressure in the gap is positive while pulling up the gel when the pressure is negative.
The asymmetric traction
breaks the symmetry of the gap thickness profile, $h(x)$, thus giving rise to a repulsive
force of hydrodynamic origin. The pressure profile and gap thickness shown here are calculated
for a thin elastic layer (\S \ref{thine}) for a dimensionless deflection 
$\eta=10$.}
\label{schematic1}
\end{center}
\end{figure}

To address some of these questions we start with the simple case of two fluid-lubricated rigid non-conforming surfaces sliding past one another at a velocity $V$ as shown in figure \ref{schematic1}.  The  viscous stresses and pressure gradient due to flow in the narrow  contact zone are  dominant.   For a Newtonian fluid, the Stokes equations (valid in the gap) are reversible in time, $t$, so that the transformation $t\to-t$ implies the transformations of the velocity $V\to-V$ and the normal force $L\to-L$.  In the vicinity of the contact region non-conforming surfaces are  symmetric which implies that  these flows are identical and therefore $L=0$. Elastohydrodynamics alters this picture qualitatively.  In front of the slider the pressure is positive and pushes down the substrate, while behind the slider the pressure is negative and pulls up the substrate.  As the solid deforms, the symmetry of the gap profile is broken leading to a normal force which pushes the cylinder away from the substrate.

\begin{figure}
\begin{center}
\includegraphics[width=7cm]{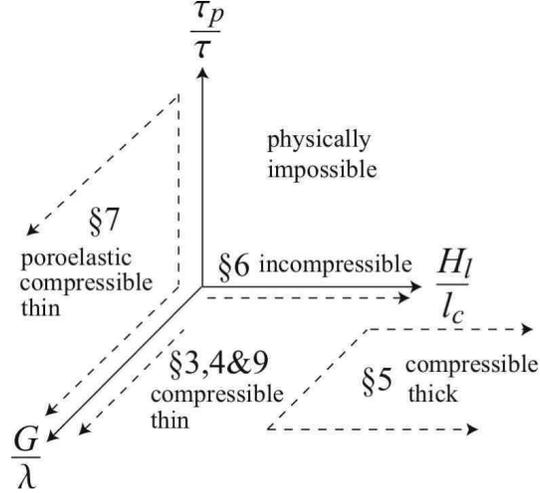}
\par
\caption{Overview of the different geometries and elastic materials considered.
$G$ and $\lambda$ are the Lam\'e coefficients of the linear elastic material, where $G/\lambda=0$ 
corresponds to an incompressible material,
$H_l$ is the depth of the elastic layer coating a rigid surface, $l_c$ is the contact length,
$\tau_p$ is the time scale over which stress relaxes in a poroelastic medium (a material composed
of an elastic solid skeleton and an interstitial viscous fluid), and $\tau = l_c/V$ is the time scale of the 
motion.
 \S \ref{thine} treats
normal-tangential coupling of non-conforming contacts coated with a thin compressible elastic layer.
\S \ref{Dsec} treats normal-tangential coupling of higher order degenerate contacts
coated with a thin compressible elastic layer.
\S \ref{6} treats normal-tangential coupling of non-conforming contacts coated
with a thick compressible elastic layer.
\S \ref{4} treats normal-tangential coupling of non-conforming contacts coated with an 
incompressible elastic layer.
\S \ref{5} treats normal-tangential coupling of non-conforming contacts coated with a thin
compressible poroelastic layer. 
\S \ref{7}  treats normal-tangential coupling of non-conforming contacts between a rigid solid and a
cylindrical shell.
\S \ref{8} treats elastohydrodynamic effects due to coating a journal bearing with a thin compressible
elastic layer.
\S \ref{10} treats elastohydrodynamic effects for 3-dimensional flows using scaling analysis.
 }
\label{summary}
\end{center}
\end{figure}

This picture applies naturally to soft  interfaces which arise either due to the properties of the material involved, as in the case of gels,  or the underlying geometry, as in the case of thin shells.
 In  \S \ref{thine} we study the normal-tangential coupling of a non-conforming contact coated with a thin compressible elastic layer. If the gap profile prior to elastic deformation is parabolic in the vicinity of the contact, the 
contact is non-conforming. However, if a parabolic description prior to deformation is insufficient we refer to the contact as conforming; {\it e.g.} the degenerate case considered in \S\ref{Dsec}. \S \ref{6} treats normal-tangential coupling of non-conforming contacts coated with a thick compressible elastic layer.
In \S \ref{4} we consider the normal-tangential coupling of non-conforming contacts coated with an 
incompressible elastic layer. In \S \ref{5} we treat the normal-tangential coupling of non-conforming contacts coated with a thin compressible poroelastic layer which describes a biphasic material composed of an elastic solid matrix and a viscous fluid (Biot 1941).  In \S \ref{7} we study the normal-tangential coupling of a non-conforming contact where one solid is rigid and the other is a deformable cylindrical shell. In \S \ref{8} we study a conforming contact: a journal bearing coated with a thin compressible elastic layer; Finally,  in \S \ref{10} we treat the elastohydrodynamics  of 3-dimensional flows using scaling analysis. Figure \ref{summary} provides an overview of the different geometries and 
elastic materials considered.

Our detailed study of a variety of seemingly distinct physical systems allows us clearly 
observe their similarities and to outline a robust set of features we expect to see in 
any soft contact. In all the cases studied the normal force = contact area $\cdot$ characteristic hydrodynamic pressure $\cdot L(\eta)$, where $L(\eta)$ is the dimensionless lift and the softness parameter   $\eta = \frac{\hbox{hydrodynamic pressure}}{\hbox{elastic stiffness}} f(\hbox{geometry}) = 
\frac{\hbox{elastic surface displacement}}{\hbox{characteristic gap thickness}}$.
Tables \ref{t.1} and \ref{t.2} summarize our results for $\eta \ll 1$. Increasing $\eta$ increases the asymmetry of the gap profile which results in a repulsive elastohydrodynamic force, {\it i.e.} in the generation of lift forces.  However, increasing $\eta$ also decreases the 
magnitude of the pressure distribution.  The competition between symmetry breaking,
which dominates for small $\eta$, and decreasing pressure, dominant at large $\eta$, 
produces an optimal combination of geometric and material parameters, $\eta_{max}$, 
that maximize the dimensionless lift, $L$.  Whether or not the normal force has a maximum depends on the control parameter: the normal force increases monotonically with the velocity, but has a maximum as a function of the effective elastic modulus of the system. This suggests that a judicious choice of material may aid in the generation of repulsive elastohydrodynamic forces thereby reducing friction and wear.

\section{Fluid lubrication theory}
\label{ltheory}

We consider a cylinder of radius $R$ moving at a velocity $V$ and
rotating with angular frequency $\omega$ and immersed 
completely in a fluid of viscosity $\mu$ as shown in figure \ref{schematic1}.
The surfaces are separated by a distance $h(x)$, the gap profile, where
the $x-$direction is parallel to the solid surface and the $z-$direction is perpendicular to it. 
We assume that the velocity and pressure field are two-dimensional
and in the region of contact we use a parabolic approximation,
valid for all non-conforming contacts, 
for the shape of the cylindrical surface in the absence of any elastic deformation.  
Then the total gap between the cylinder and the solid is given by
\begin{equation}
\label{1}
h(x)= h_0(1 + \frac{x^2}{2h_0R}) + H(x),
\end{equation}
with $H$ begin the additional elastic deformation and 
$h_0$ the characteristic gap thickness in the absence of solid deformation.  
The size of the contact zone 
$\sqrt{2h_0R}$, 
characterizes the horizontal size over which the lubrication forces are important. 
consistent with the parabolic approximation in (\ref{1}).
If $h_0\ll R$, the gap Reynolds number
Re$_{\hbox{g}} = \frac{\rho V^2/l}{\mu V/h_0^2} \sim \frac{\rho V h_0^{3/2}}{\mu R^{1/2}}
\ll \frac{\rho V R}{\mu} = $ Re, the nominal Reynolds number.  
Then, if Re$_{\hbox{g}}\ll1$ 
we can neglect the inertial terms and use the lubrication approximation (Reynolds 1886) 
to describe the hydrodynamics.
For a 2-dimensional velocity field ${\bf v} = (v_x(x,z),v_z(x,z))$ and a pressure field $p(x,z)$
the fluid stress tensor is 
\begin{equation}
\label{fst1}
\boldsymbol{\sigma}_f = \mu (\nabla {\bf v} + \nabla {\bf v}^T) - p {\bf I}.
\end{equation}
Stress balance in the fluid, $\nabla \cdot \boldsymbol{\sigma_f} = 0$, yields
\begin{eqnarray}
0=\partial_z p, \nonumber \\
\label{S1}
0=-\partial_xp + \mu   \partial_{zz} v_x.
\end{eqnarray}
Mass conservation implies
\begin{equation}
\label{cont}
0=\partial_xv_x+\partial_zv_z.
\end{equation}
The associated boundary conditions are
\begin{eqnarray}
v_x|_{z=-H} = -V,~&~v_x|_{z=h_0+ \frac{x^2}{2R}}=-\omega R,\nonumber \\
v_z|_{z=-H}= V \partial_x H,~&~v_z|_{z=h_0+ \frac{x^2}{2R}}=-x  \omega, \nonumber \\
\label{Sbc}
p|_{x\to\infty} = 0,& p|_{x\to-\infty}=0,
\end{eqnarray}
where we have chosen to work in a reference frame translating with the cylinder.
We make the variables dimensionless with the following definitions
\begin{eqnarray}
x=\sqrt{2h_0R} \, x',~~~z=h_0z',~~~p=p_0\,p'=\frac{\sqrt{2R}\mu (V-\omega R)}{h_0^{3/2}}p', \nonumber \\
\boldsymbol{\sigma}_f=p_0 \, \boldsymbol{\sigma}_f'=\frac{\sqrt{2R}\mu (V-\omega R)}{h_0^{3/2}}\boldsymbol{\sigma}_f',~~~
h = h_0h',~~~H=H_0 \,H',
\nonumber \\
\label{fscale}
 v_z=(V-\omega R)\sqrt{\frac{h_0}{2R}}v'_z,~~~v_x=(V-\omega R)v'_x.
\end{eqnarray}
Here, $H_0(\frac{p_0}{E'},\frac{l_2}{l_1},\frac{l_3}{l_1},...)$ is the characteristic scale of the deflection, 
where $E'$ is the effective elastic modulus of the medium and $l_i$ are the length scales of the system.
The pressure scaling follows from (\ref{S1}) and the fact that $x\sim\sqrt{h_0R}$.
Then, after dropping the primes the dimensionless versions of equations (\ref{S1})-(\ref{Sbc}) are
\begin{eqnarray}
\label{solve}
0=\partial_z p, \nonumber \\
0=-\partial_xp + \partial_{zz} v_x, \nonumber \\
0=\partial_xv_x+\partial_zv_z, \nonumber \\
v_x|_{z=-\eta H} = \frac{-1}{1-\omega'},~~~v_x|_{z=1+x^2}=\frac{-\omega'}{1-\omega'},\nonumber \\
v_z|_{z=-\eta H}=\frac{\eta\, \partial_x H}{1-\omega'},~~~v_z|_{z=1+x^2}=\frac{-2x\omega'}{1-\omega'}. \nonumber \\
\label{Ssystem}
p|_{x\to\infty} = 0, ~~~ p|_{x\to-\infty}=0
\end{eqnarray}
Here $\omega' = \omega R/V$ characterizes the ratio of rolling to sliding,
the softness parameter
$\eta= H_0/h_0$ characterizes the scale of the elastic deformation relative to the gap thickness,
which is related to the compliance of the elastic material.
The dimensionless version of the fluid stress tensor (\ref{fst1}) is 
\begin{equation}
\label{fst}
\boldsymbol{\sigma}_f =
\left( \begin{array}{cc}
- p + 2 \varepsilon^2 \partial_xv_x & 
\varepsilon \partial_z v_x + \varepsilon^3 \partial_x v_z \\
\varepsilon \partial_z v_x + \varepsilon^3 \partial_x v_z & 
- p + 2 \varepsilon^2 \partial_zv_z
\end{array}
\right).
\end{equation}
where $\varepsilon = \sqrt\frac{h_0}{2R}$.
Solving (\ref{Ssystem}) gives the Reynolds equation (Batchelor 1967)
\begin{equation}
\label{Req}
0=\partial_x(6h+h^3\partial_xp), 
\end{equation}
subject to
\begin{equation}
\label{Reqbc}
p(\infty)=p(-\infty)=0.
\end{equation}
Note that $\omega'$ is scaled away.  Here, $h$ is the gap profile given by (\ref{1}) in dimensionless terms
\begin{equation}
\label{h(x)}
h(x)=1+x^2+\eta H(x).
\end{equation}
To close the system we need to determine $\eta H(x)$, the elastic response to the hydrodynamic forces.
This depends on the detailed geometry and constitutive behavior of the cylindrical contact. 
In the following sections we explore various configurations that allow us to explicitly calculate $\eta H$, thus 
allowing us to calculate the normal force on the cylinder 
\begin{equation}
L = \int_{\hbox{contact area}} p\,dA,
\end{equation}
and determine the elastohydrodynamic tangential-normal coupling.  
We note that when $\eta=0$, the contact is symmetric (\ref{h(x)}) so that the form of (\ref{Req}) 
implies that $p(-x)=-p(x)$ and $L=0$.

\section{Elastic 'lubrication' theory: deformation of a thin elastic layer}
\label{thine}

\begin{figure}
\begin{center}
\includegraphics[width=7cm]{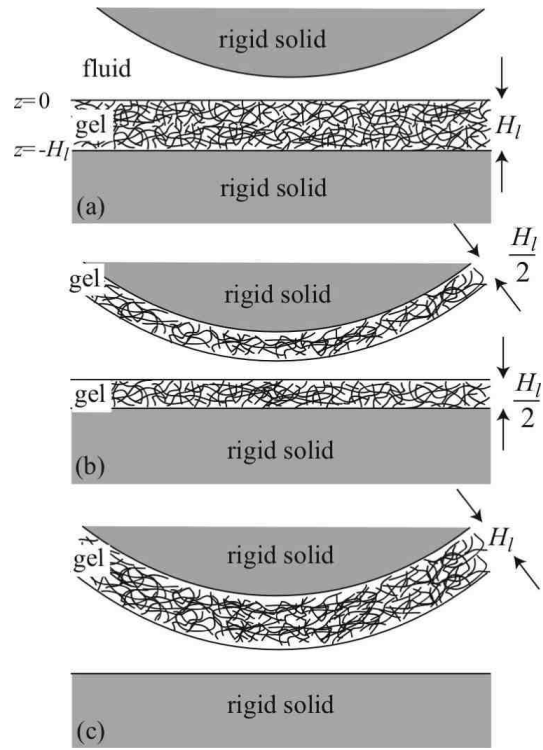}
\par
\caption{Schematic diagrams of mathematically identical configurations. Configuration (a) is treated 
in the text.}
\label{cases}
\end{center}
\end{figure}

In our first case, we consider a thin elastic layer of thickness $H_l \ll l_c$ coating the
cylinder or the rigid wall or both, all of which are mathematically equivalent  (figure \ref{cases}).  
We first turn our attention to determining the surface deflection of the layer for an arbitrary applied
traction.  Throughout the analysis we assume that the surface deflection $H_0\ll H_l$ so that 
a linear elastic theory suffices to describe the material response. 
The stress tensor, $\boldsymbol{\sigma}_s$, 
for a linearly elastic isotropic material with Lam\'e coefficients $G$ and $\lambda$ is 
\begin{equation}
\label{stressT}
\boldsymbol{\sigma}_s = G (\nabla {\bf u} + \nabla{\bf u}^T) + \lambda\nabla \cdot{\bf u}\, {\bf I},
\end{equation}
where ${\bf u}=(u_x,u_z)$ is the displacement field, and ${\bf I}$ is the identity tensor. 
Stress balance in the solid implies
\begin{equation}
\label{fbalance}
\nabla\cdot\boldsymbol{\sigma}_s=0.
\end{equation}
We make the equations dimensionless using
\begin{eqnarray}
z = H_l\,z',  ~~~ x =  \sqrt{2 h_0R}\,x', ~~~ u_x = h_0 u_x', ~~~ u_z = h_0 u_z'  ~~~ \boldsymbol{\sigma}_s = p_0 \boldsymbol{\sigma}_s'.
\end{eqnarray}
We note that the length scale in the $z-$direction is the depth of the layer $H_l$; 
the length scale in the $x-$direction is the length scale of the hydrodynamic contact zone, 
$l_c = \sqrt{2 h_0R}$; the displacements $u_x$ and $u_z$ have been scaled with 
the characteristic gap thickness $h_0$; and the 
stress has been scaled using the hydrodynamic pressure scale following (\ref{fscale}).  
We take the thickness of the solid layer to be small compared to the length scale of the 
contact zone with $\zeta = \frac{H_l}{\sqrt{2 h_0R}}\ll1$, and restrict our attention to 
compressible elastic materials, where 
$G \sim \lambda$. Then, after dropping primes, the dimensionless 2-dimensional form of 
the stress tensor (\ref{stressT}) is 
\begin{equation}
\boldsymbol{\sigma}_s=
\frac{1}{\eta} \left( \begin{array}{cc}
 \frac{\lambda}{2G+\lambda} \partial_z u_z + \zeta \partial_x u_x & 
 \frac{G}{2G+\lambda} \partial_z u_x + \zeta \frac{G}{2G+\lambda} \partial_x u_z \\
  \frac{G}{2G+\lambda} \partial_z u_x + \zeta \frac{G}{2G+\lambda} \partial_x u_z & 
  \partial_z u_z + \zeta \frac{\lambda}{2G+\lambda} \partial_x u_x
\end{array}
\right)
\end{equation}
Here  
\begin{equation}
\label{eta1}
\eta =\frac{p_0}{2G+\lambda} \frac{H_l}{h_0} =
\sqrt{2} \frac{\mu (V-\omega R)}{2G + \lambda} \frac{ H_l R^{1/2}}{h_0^{5/2}}
\end{equation}
is the softness parameter, a dimensionless number 
governing the relative size of the surface deflection to the undeformed gap thickness.  
Stress balance (\ref{fbalance}) yields 
\begin{eqnarray}
\partial_{zz} u_x + \zeta (1+\frac{\lambda}{G}) \partial_{xz} u_z + \zeta^2 (2+\frac{\lambda}{G}) \partial_{xx}u_x=0,
\nonumber \\
\label{ZF}
\partial_{zz}u_z + \zeta \frac{G+\lambda}{2G + \lambda} \partial_{xz}u_x + \zeta^2 \frac{G}{2G + \lambda} \partial_{xx}u_z=0,
\end{eqnarray}
so that to $O(\zeta)$ the leading order balance is
\begin{eqnarray}
\partial_{zz} u_x = 0, \nonumber \\ 
\label{vert}
\partial_{zz} u_z  = 0.
\end{eqnarray}
The normal unit vector to the soft interface 
is ${\bf n} = (-\partial_x u_z|_{z=0},1)$, which in 
dimensionless form is 
\begin{equation}
{\bf n} = ( -\varepsilon \partial_x u_z|_{z=0},1),
\end{equation}
where $\varepsilon = \sqrt\frac{h_0}{2R}$.
The balance of normal traction on the solid-fluid interface yields 
\begin{equation}
\boldsymbol{\sigma}_f\cdot{\bf n}|_{z=0} = \boldsymbol{\sigma}_s\cdot{\bf n}|_{z=0},
\end{equation}
so that 
\begin{eqnarray}
\label{uzsurf}
\partial_z u_x|_{z=0} =0, ~~~
\partial_z u_z|_{z=0}= - \eta \, p.
\end{eqnarray}
At the interface between the soft film and the rigid substrate, the no slip condition yields
\begin{equation}
\label{vertbc}
u_z(x,-1)=0, ~~~ u_x(x,-1)=0.
\end{equation}
Solving  (\ref{vert}), (\ref{uzsurf}) and (\ref{vertbc}) gives us the displacement of the solid-fluid interface
\begin{equation}
\label{locald}
u_x(x,0) = 0, ~~~ u_z(x,0) = -\eta\,p = -\eta H(x).
\end{equation}
This linear relationship between the normal displacement and fluid pressure 
is known as the Winkler or 'mattress' elastic foundation
model (Johnson 1985).  In light of (\ref{locald}) we may write the gap profile (\ref{h(x)}) as
\begin{equation}
\label{limit}
h=1+x^2+ \eta \, p.
\end{equation}
Equations (\ref{Req}), (\ref{Reqbc}) and (\ref{limit}) form a closed system for the elastohydrodynamic response 
of a thin elastic layer coating a rigid cylinder.  
We note that Lighthill (1968) found a similar set of equations while 
studying the flow of a red blood cell through a capillary.
However, his model's axisymmetry proscribed the existence of a force normal to the flow.  

\begin{figure}
\begin{center}
\includegraphics[width=8cm]{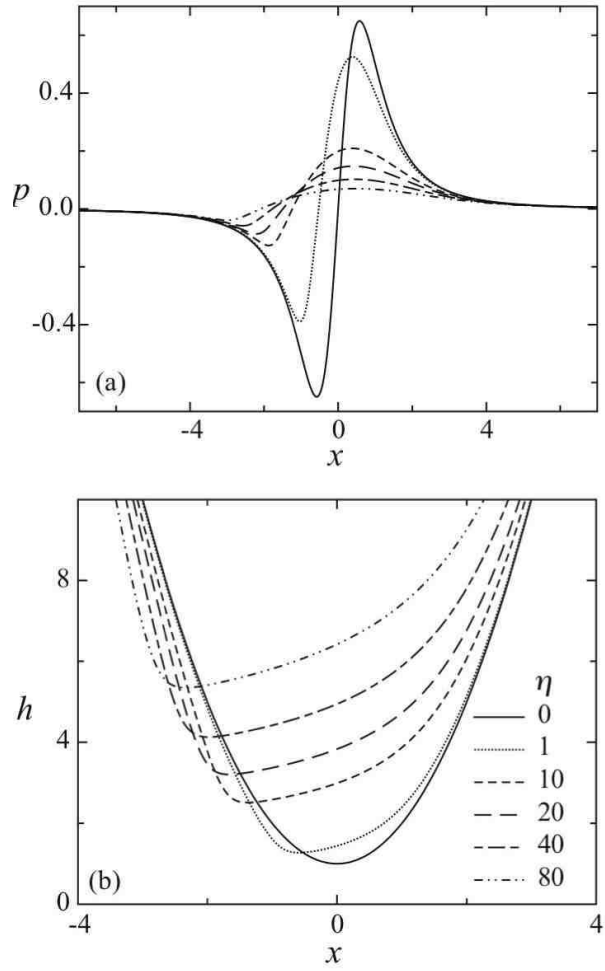}
\par
\caption{(a) Pressure $p(x)$ as a function of $\eta$. (b) Gap thickness profile, $h(x) = 1 + x^2 + \eta p$
as a function of $\eta$.  
The initially parabolic gap thickness profile is broken 
and the maximum value of the pressure decreases as $\eta$ increases.}
\label{bbb}
\end{center}
\end{figure}

\begin{figure}
\begin{center}
\includegraphics[width=7cm]{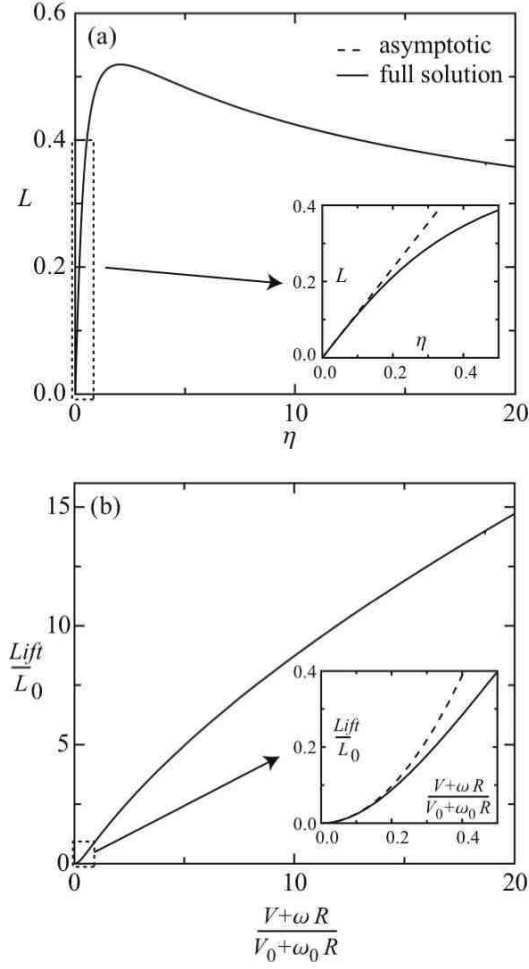}
\par
\caption{(a) Dimensionless lift per unit length, $L$,  plotted against $\eta$, the softness parameter.  
$L$ has a maximum at $\eta=2.06$ which is the 
result of a competition between symmetry breaking (dominant for $\eta \ll 1$) 
and decreasing pressure (dominant for $\eta \gg 1$) due to increasing 
the gap thickness. For small $\eta$ asymptotic analysis yields $L=\frac{3\pi}{8}\eta$, which 
matches the numerical solution. 
(b) The dimensional lift force, $Lift$, is quadratic in the velocity for small velocities
while being roughly linear for large velocities. $V_0+\omega_0R = \frac{h_0^{5/2}(2G+\lambda)}{\sqrt{2R}H_l\mu }$ is the velocity and rate of rotation at which $\eta=1$, and  
$L_0$ is corresponding lift.  
}
\label{elastic}
\end{center}
\end{figure}

When $\eta \ll 1$ we can employ a perturbation analysis to find
the lift force experienced by the cylinder.  
We use an expansion of the form
\begin{equation}
\label{expansion}
p = p^{(0)} + \eta p^{(1)} + O(\eta^2),~~~h = h^{(0)} + \eta h^{(1)} + O(\eta^2),
\end{equation}
to find 
\begin{eqnarray}
\label{eta00}
\eta^0 : ~\partial_x \{ 6 h^{(0)}+ [h^{(0)}]^3\partial_xp^{(0)} \}=0,\\
\label{eta11}
\eta^1 :~\partial_x \{ 6  h^{(1)} + 3  [h^{(0)}]^2 h^{(1)} \partial_xp^{(0)} + 
  [h^{(0)}]^3\partial_xp^{(1)} \}=0,
\end{eqnarray}
where 
\begin{equation}
 h^{(0)} = 1+x^2, ~~~  h^{(1)} =  p^{(0)},
 \end{equation}
subject to the boundary conditions 
\begin{equation}
\label{pexpbc}
p^{(0)}(\infty)=p^{(0)}(-\infty)=p^{(1)}(\infty)=p^{(1)}(-\infty)=0.
\end{equation}
Solving (\ref{eta00})-(\ref{pexpbc}) yields
\begin{equation}
\label{p00}
p = \frac{2 x}{(1+ x^2)^2} + \eta  \frac{3(3-5x^2)}{5(1+x^2)^5}+ O(\eta^2),
\end{equation}
so that 
\begin{eqnarray}
 L  = \int p\, dx = \frac{3\pi}{8} \eta.
\end{eqnarray}
In dimensional form the lift force per unit length is
\begin{equation}
\label{dimlift}
L= \frac{3\sqrt{2}\pi}{8}\frac{p_0^2}{2G+\lambda}\frac{H_l\sqrt{R}}{\sqrt{h_0}} =\frac{3\sqrt{2}\pi}{4}\frac{\mu^2(V-\omega R)^2}{2G+\lambda}\frac{H_lR^{3/2}}{h_0^{7/2}},
\end{equation}
as reported in Skotheim \& Mahadevan (2004b). 
The same scaling was found by Sekimoto \& Leibler (1993), but with a different prefactor
owing to a typographical error.
When $\eta = O(1)$, the system (\ref{Req}), (\ref{Reqbc}) and (\ref{limit}) is solved numerically using a
continuation method (Doedel {\it et al.} 2004) with $\eta$ as the continuation parameter.
In figure \ref{bbb} we show the pressure distribution $p(x)$, and gap $h(x)$ as a function of $\eta$.
For $\eta \ll 1$, $p(-\frac{9\eta}{10})=0$.  
As $\eta$ increases, $h$
increases and the asymmetric gap profile begins to resemble that of a tilted slider bearing, which is
well known to generate lift forces.  
However, an increase in the gap thickness also decreases the
peak pressure $\sim \mu V R^{1/2}/h_0^{3/2}$ (see figure \ref{bbb}a).
The competition between symmetry breaking, dominant for $\eta \lesssim 1$, and decreasing pressure, dominant for $\eta \gtrsim 1$, produces a maximum scaled lift force 
at $\eta = 2.06$.  In dimensional terms, this implies that the lift as a function of the 
effective modulus $2G + \lambda$ will have a maximum, however, the lift as a function of 
the relative motion between the two surfaces $V-\omega R$ increases monotonically 
(see figure \ref{elastic}b).  
In fact, (\ref{dimlift}) shows that the dimensional lift increases as $(V-\omega R)^2$ for $\eta \ll 1$.

\section{Degenerate contact}
\label{Dsec}

\begin{figure}
\begin{center}
\includegraphics[width=10cm]{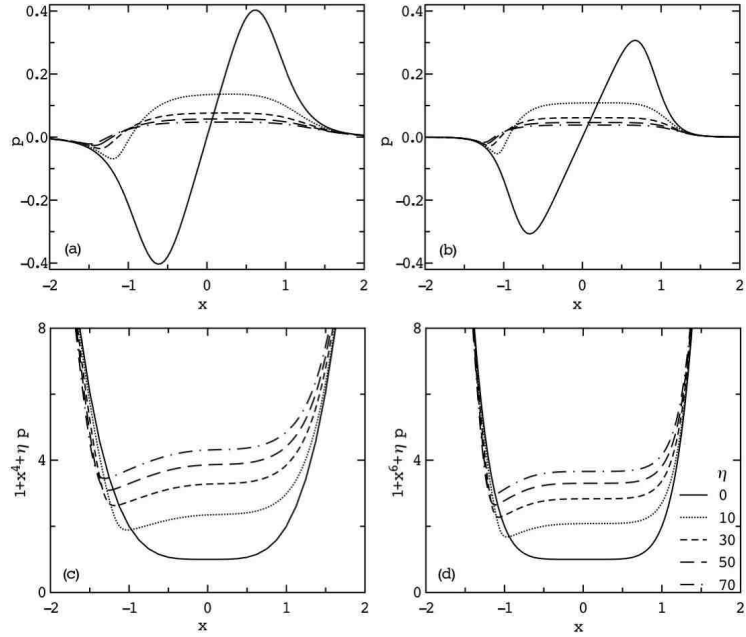}
\par
\caption{Pressure distribution and gap thickness profile for degenerate contacts corresponding to 
a gap thickness profile of $h=1+x^{2n}+\eta p$ with $n=2$
(a), (c), and $n=3$ (b), (d).}
\label{hhh}
\end{center}
\end{figure}

\begin{figure}
\begin{center}
\includegraphics[width=10cm]{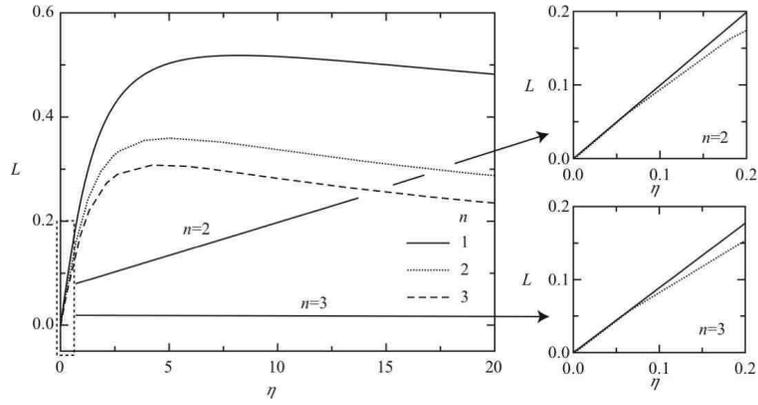}
\par
\caption{Dimensionless lift for $h=1+x^{2n}+\eta p$ where $n=1,2,3$.    
The curves are similar, however, they can not be rescaled to a universal curve.}
\label{ccc}
\end{center}
\end{figure}

In this section we consider the case where the parabolic approximation in the vicinity of the contact breaks
down.  
Since rotation changes the nature of the contact region for such interfaces, we consider only 
a purely sliding motion with $\omega=0$. 
We assume that the gap thickness is described by 
\begin{equation}
\label{hdeg}
h = h_0(1 + \frac{x^{2n}}{h_0 R^{2n-1}}) + H(x),
\end{equation}
where $n=2,3,...$ characterizes the geometric nature of the contact 
and the contact length is $l_c \sim (h_0R^{2n-1})^{1/2n}$.  
We note that we always focus on symmetric contacts.
We make the variables dimensionless using the following scalings
\begin{eqnarray}
h = h_0 h',~~~
x = l^*x' = h_0^\frac{1}{2n}R^{1-\frac{1}{2n}}x', \nonumber \\
p = p^*p'=  \mu V \frac{R^{1-\frac{1}{2n}}}{h_0^{2-\frac{1}{2n}}}\,p'.
\end{eqnarray}
In \S \ref{thine} we have seen that for a thin compressible soft layer 
the pressure and and surface deflection
can be linearly related by (\ref{locald}) so that the scale of the deflection is 
$h_0 \eta = \frac{p_0H_l}{2G + \lambda}$.  To find the scale of the deflection for the degenerate 
contact described by (\ref{hdeg}), we replace $p_0$ with the appropriate pressure scale $p^*$ so that 
\begin{equation}
H= \frac{p^*}{2G+\lambda} H_l \, H' = \frac{\mu V}{2G + \lambda} \frac{H_l R^{1 - \frac{1}{2n}}}{h_0^{2-\frac{1}{2n}}}, 
\end{equation}
and the size of the deformation relative to the gap size is governed by the dimensionless group
\begin{equation}
\label{etadeg}
\eta =   \frac{p^*}{2G+\lambda} \frac{H_l}{h_0}
=\frac{\mu V}{2G+\lambda} \frac{H_lR^{1-\frac{1}{2n}}}{h_0^{3-\frac{1}{2n}}}.
\end{equation}
Then the  dimensionless version of the gap thickness profile (\ref{hdeg}) is
\begin{equation}
\label{ndegh}
h = 1 + x^{2n} + \eta \, p.
\end{equation}
As in \S\ref{thine}, 
we employ a perturbation expansion for the pressure field in the parameter $\eta$ ($\ll1$) 
to solve (\ref{Req}), (\ref{Reqbc}) and (\ref{ndegh}) 
and find the 
pressure field and the lift for small $\eta$.  This yields $p = p_0 + \eta\,p_1$ and 
gives the following dimensionless result
\begin{equation}
L_{n=2} = \frac{351\pi}{784\sqrt{2}}\eta, ~~~ L_{n=3} = 0.8859\, \eta,
\end{equation}
where we have not shown the pressure distribution due to its unwieldy size.
In dimensional terms, the normal force reads as
\begin{eqnarray}
\label{liftf}
L_{n=2} 
=\frac{351\pi}{784\sqrt{2}} \frac{p^{*2}}{2G + \lambda} \frac{H_ll^*}{h_0}
= \frac{351\pi}{784\sqrt{2}}\frac{\mu^2 V^2 H_lR^{9/4}}{(2G + \lambda)h_0^{17/4}}, \\
L_{n=3}= 0.8859\frac{p^{*2}}{2G + \lambda} \frac{H_ll^*}{h_0}
 = 0.8859 \frac{\mu^2 V^2 H_lR^{5/2}}{(2G + \lambda)h_0^{9/2}}.
\end{eqnarray}
The results for $n=2$ and $n=3$ are shown in figures \ref{hhh} and \ref{ccc}.
One noteworthy feature of a degenerate contact is that the torque experienced by the slider 
arises from the fluid pressure rather that the shear force because the normal to the surface
no longer passes through the center of the object as it would for a cylinder or sphere.  
The ratio of the torque due to shear to 
the torque due to the pressure is 
\begin{equation}
\frac{\hbox{shear torque}}{\hbox{pressure torque}}\sim \frac{\mu V R/h_0}{p\,l_c} \sim (\frac{h_0}{R})^{1-\frac{1}{n}}\ll1.
\end{equation}
Hence, the dominant contribution to the torque is due to the pressure and 
\begin{equation}
\Gamma = \int p\,x\,dA.
\end{equation}
For $\eta\ll1$, $p = p_0 + O(\eta)$ so that
\begin{eqnarray}
n=2\,:~~~ p_0 = \frac{6 x}{7 (1+x^4)^2}, ~~~ \Gamma = \frac{3\,\pi}{14\sqrt{2}}, \nonumber \\
n=3\,:~~~ p_0 =  \frac{6 x}{11 (1+x^6)^2}, ~~~ \Gamma = \frac{\pi}{11} , \nonumber \\
n = m\,:~~~p_0 = \frac{6 x}{(4m-1)(1+x^{2m})^2}, ~~~ \Gamma = \frac{3 \pi (2m-3) \csc\frac{3\pi}{2 m}}{m^2(4m-1)}.
\end{eqnarray} 

\section{Soft slider}
\label{6}

\begin{figure}
\begin{center}
\includegraphics[width=10cm]{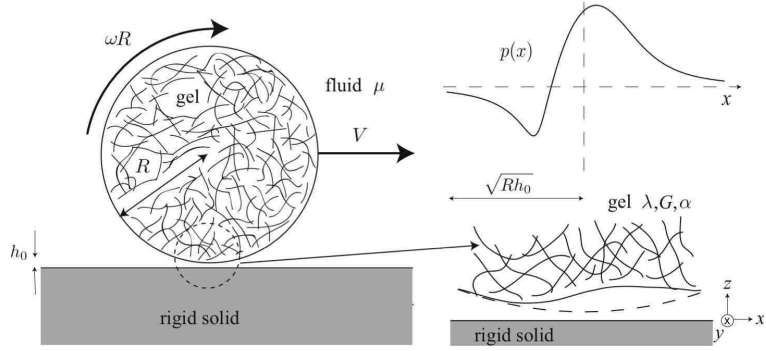}
\par
\caption{Schematic diagram of a gel cylinder moving through a liquid over a
rigid solid substrate.  The asymmetric pressure distribution pushes on the gel when
the fluid pressure in the gap is positive while pulling on the gel when the pressure is negative.
This breaks the symmetry of the gap thickness profile, $h(x)$, and gives rise to a repulsive
force of elastohydrodynamic origin. The dashed line in the lower right hand denotes the undeformed location of the gel cylinder.}
\label{schematic}
\end{center}
\end{figure}

\begin{figure}
\begin{center}
\includegraphics[width=8cm]{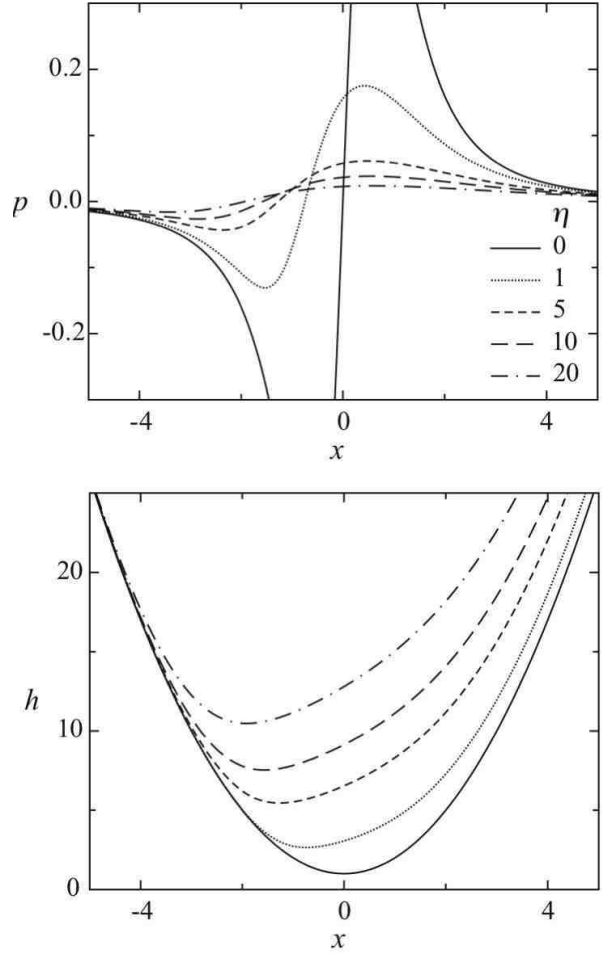}
\par
\caption{Gap thickness $h$ and pressure $p$ as a function of $\eta$ for a
soft cylindrical gel slider.  We note that while the 
pressure distribution is localized to the region near the point of closest contact, the change in 
gap thickness is spread out due to the 
logarithmic nature of the Green's function of a line contact: 
$h =1+x^2+ \eta \int dx' p(x') \log[\frac{Y}{(x-x')^2}]$. }
\label{softh}
\end{center}
\end{figure}

\begin{figure}
\begin{center}
\includegraphics[width=8cm]{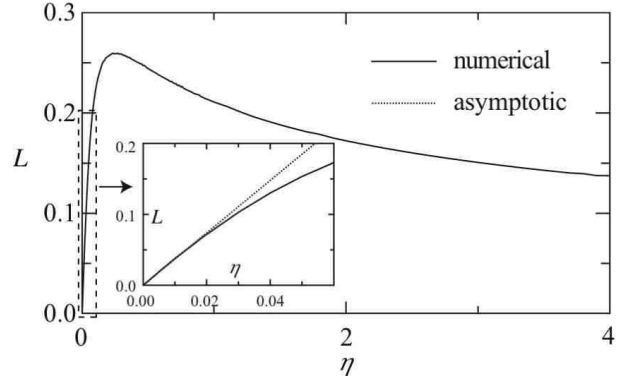}
\par
\caption{Dimensionless lift as a function of $\eta$ a measure of the increase in gap thickness 
for a soft cylindrical gel slider.  For $\eta \ll 1$, $L = \frac{3\pi^2}{8}\eta$.}
\label{softlift}
\end{center}
\end{figure}

To contrast our result for a thin layer with that for a soft slider,
we consider the case where the entire cylindrical 
slider of length $l$ and radius $R$ 
is made of a soft material with Lam\'e coefficients $G$ and $\lambda$. 
Equivalently, we could have a rigid slider moving above a soft semi-infinite half space.
Since the deformation
is no longer locally determined by the pressure 
we use a Green's function approach to determine the response to the hydrodynamic 
pressure.  
Following Davis, Serayssol \& Hinch (1986), we use the Green's function for a point force
on a half space since the scale of the contact zone, $l_c\sim\sqrt{h_0R} \ll R$, the cylinder radius.
Following Landau \& Lifshitz (1970) we write the 
deformation at the surface due to a pressure field $p(x,y)$ as
\begin{equation}
\label{gfcn}
H(x,y) = \frac{\lambda + 2G}{4\pi G(\lambda+G)} \int \frac{p(x',y')dx'\,dy'}{\sqrt{(x'-x)^2+(y'-y)^2}},
\end{equation}
with $x,~y$ as defined in figure \ref{schematic}.
Neglecting end effects so the pressure is a function of $x$ only, we
integrate (\ref{gfcn}) over $y'$ to get
\begin{eqnarray}
H(x) = &   \frac{\lambda + 2G}{4\pi G(\lambda+G)} \int_{-\infty}^\infty \left[ \,p(x') \int_{-l}^l \frac{dy'}{\sqrt{(x'-x)^2+(y'-y)^2}}\right]dx'  \nonumber \\
\label{dh2}
=  & \frac{\lambda + 2G}{4\pi G(\lambda+G)} \int_{-\infty}^\infty 
\log\left[\frac{4(l^2-y^2)}{(x-x')^2}\right] p(x') dx' ,
\end{eqnarray}
To make the equations dimensionless we employ the following scalings
\begin{eqnarray}
x = \sqrt{2h_0R} \,x',~~~~~ y = l\,y', ~~~~~ h = h_0\,h', \nonumber \\
H = h_0\,H', ~~ p = p_0 p' = \frac{\sqrt{2R}\mu (V-\omega R)}{h_0^{3/2}}p',
\end{eqnarray}
so that the dimensionless gap thickness (\ref{h(x)}) 
now reads
\begin{equation}
\label{DH}
h = 1+x^2+ \eta \int_{-\infty}^\infty dx' p(x') \log[\frac{Y}{(x-x')^2}]
\end{equation}
where 
\begin{equation}
\label{softeta1}
\eta = \frac{1}{2\pi}\frac{\mu (V-\omega R)\,(\lambda + 2G)}{G(\lambda+G)}\frac{R}{h_0^2}, ~~ 
Y= \frac{2l^2}{Rh_0}(1-y^2).
\end{equation}
Comparing with our softness parameter for a thin section we see that 
\begin{equation}
\frac{\eta_{\hbox{thin layer}}}{\eta_{\hbox{soft slider}}} = 2\pi\sqrt{2} \frac{G(G + \lambda)}{(2G+\lambda)^2} \frac{H_l}{\sqrt{h_0R}} \sim \frac{H_l}{\sqrt{h_0R}} \ll 1,
\end{equation} 
{\it i.e.} a thin layer is stiffer than a half space made from the same material by the geometric factor
$\frac{\sqrt{h_0R}}{H_l}$.
For small $\eta$ we write $p=p_0+\eta p_1$  where $p_0=\frac{2x}{(1+x^2)^2}$ as in (\ref{p00}).
Substituting into (\ref{DH}) yields
\begin{equation}
h=1+x^2+\eta \frac{2\pi\,x}{1+x^2} + O(\eta^2).
\end{equation}
To order $\eta$, (\ref{Req}) and (\ref{Reqbc}) yield the equations for the perturbation pressure, $p_1$:
\begin{eqnarray}
\eta:~~ 0=\partial_x[(1+x^2)^3 \partial_x p_1 + \frac{24\pi\,x\,(x^2-1)}{(1+x^2)^2}], \nonumber \\
p_1(-\infty)=p_1(\infty) = 0,
\end{eqnarray}
which has the solution
\begin{equation}
p_1 = \frac{2\pi(2x^2-1)}{(1+x^2)^4}
\end{equation}
Hence the dimensionless lift force is
\begin{equation}
L = \eta \int p_1\,dx  = \eta \int \frac{2\pi(2x^2-1)}{(1+x^2)^4}dx = \frac{3\pi^2}{8}\eta.
\end{equation}
In dimensional terms the lift force is 
\begin{equation}
L = \frac{3\pi}{8} \frac{\mu^2 (V-\omega R)^2(\lambda + 2G)}{ G(\lambda+G)}\frac{R^2}{h_0^3}.
\end{equation}
Comparing this expression with that for the case of a thin elastic layer, equation (\ref{dimlift}),
we see that confinement acts to reduce deformation and hence reduce 
the lift in the small deflection, $\eta\ll 1$, regime.  
When $\eta = O(1)$ we solve (\ref{Req}), (\ref{Reqbc}) and (\ref{DH}) for $p(x)$
using an iterative procedure.  
First, we guess an
initial gap profile $h_{old}$ and use a  
shooting algorithm to calculate the pressure distribution.
The new pressure distribution is then used in equation (\ref{DH}) with $Y=1000$ (corresponding 
to a very long cylinder) to calculate a new gap profile,
$h_{new}$.
If $\int_{-10}^{10} (h_{old} - h_{new})^2 dx < 10^{-6}$ 
the calculation is stopped, else we set $h_{old}=h_{new}$ and iterate.  
The results are shown in figures \ref{softh} and \ref{softlift}, and not surprisingly they have the
same qualitative features discussed previously, {\it i.e.} for $\eta \ll 1$, $L\sim\eta$; $L$ shows a
maximum at $\eta=0.25$ and decreases when $\eta > 0.25$.
The reasons for this are the same as before, {\it i.e.} the competing effects of an increase in the gap thickness and the increased asymmetry of the contact zone.  

\section{Incompressible layer}
\label{4}

\begin{figure}
\begin{center}
\includegraphics[width=10cm]{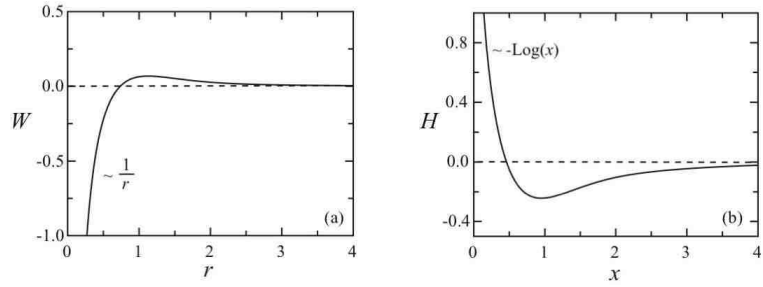}
\par
\caption{Green's function for a point force (a) and a line load (b) acting on an incompressible
layer of dimensionless thickness $\zeta = \frac{H_l}{l_c}=1$.}
\label{icgf}
\end{center}
\end{figure}

\begin{figure}
\begin{center}
\includegraphics[width=10cm]{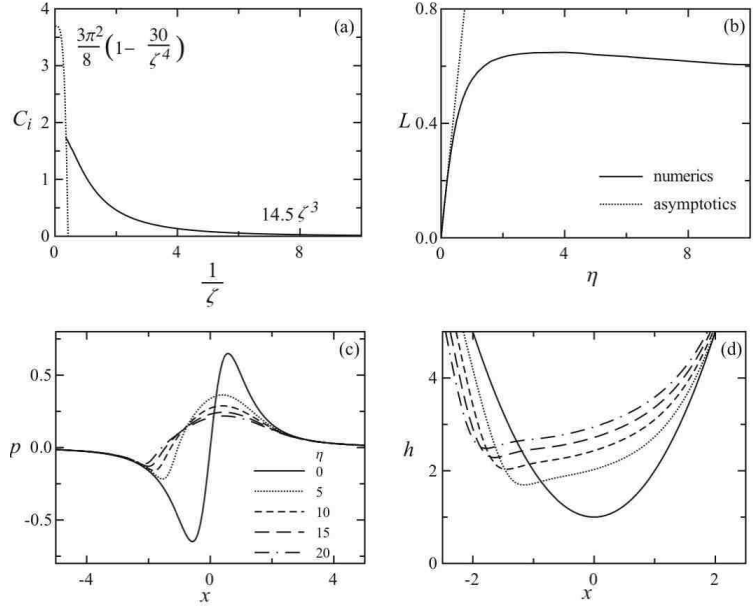}
\par
\caption{ 
For $\eta \ll 1$, $L=C_i(\zeta) \eta$
where $C_i(\zeta)$ is shown in (a).
(b) shows the dimensionless lift per unit length, $L$, as 
a function of $\eta$ for $\zeta = \sqrt{h_0 R}/H_l = 1$.   
(c) and (d) show pressure, $p$, and gap thickness, $h$, 
as a function of $\eta$.
As the thickness of the layer decreases the presence 
of the undeformed
substrate below is increasingly felt and the layer stiffens.  
In the linear regime a stiffer layer results
in a smaller deformation and concomitant decrease in lift.
}
\label{icf}
\end{center}
\end{figure}

In contrast to compressible layers, an incompressible layer ({\it e.g.} one made of an elastomer)
can deform only via shear.
For thin layers, incompressibility 
leads to a geometric stiffening that qualitatively changes the nature of 
the elastohydrodynamic problem (Johnson 1985).
To address this problem in the most general case, we use a Green's function approach.
The constitutive behavior for an incompressible linearly elastic solid is 
\begin{equation}
\boldsymbol{\sigma} = G (\nabla {\bf u} + \nabla {\bf u}^T)-p_s {\bf I},
\end{equation}
where ${\bf u} = (u_x,u_y,u_z)$ is the displacement and $p_s$ is the pressure 
in the solid.
Mechanical equilibrium in the solid implies $\nabla\cdot\boldsymbol{\sigma}=0$, {\it i.e.}
\begin{eqnarray}
0 = -\partial_x p_s + G \nabla^2 u_x, \nonumber \\
0 = -\partial_y p_s + G \nabla^2 u_y, \nonumber \\
\label{inc1}
0 = -\partial_z p_s + G \nabla^2 u_z.
\end{eqnarray}
Incompressibility of the solid implies
\begin{equation}
\label{inc2}
0 = \nabla\cdot{\bf u} = \partial_x u_x + \partial_y u_y + \partial_z u_z.
\end{equation}
For the Green's function associated with a point force,
$\sigma_{zz}|_{z=0} = - f \delta(x)\delta(y)$ where $\delta$ is a delta function, 
the boundary conditions are 
\begin{eqnarray}
&u_x = u_y = u_z = 0 ~~ &\hbox{at} ~~z = -H_l, \nonumber \\
&\sigma_{xz}=\sigma_{xy}=0 ~~ &\hbox{at}~~ z = 0, \nonumber \\
\label{inc3}
&\sigma_{zz} = - f \delta(x)\delta(y)  ~~&\hbox{at} ~~z = 0.
\end{eqnarray}
We solve the boundary value problem (\ref{inc1})-(\ref{inc3}) by using a 2-D  
Fourier transforms defined as
\begin{equation}
u_x = \int_{-\infty}^\infty  \int_{-\infty}^\infty \hat{u}_x(k_x,k_y,z)e^{-i(k_xx+k_yy)}dk_x\,dk_y,~...
\end{equation}
Then equations (\ref{inc1})-(\ref{inc3}) in Fourier space are
\begin{eqnarray}
0 = ik_x\hat p_s + G (-k_x^2 \hat u_x - k_y^2\hat u_x + \partial_{zz}\hat u_x), \nonumber \\
0 = ik_y\hat p_s + G (-k_x^2 \hat u_y - k_y^2\hat u_y + \partial_{zz}\hat u_y), \nonumber \\
0 = -\partial_z \hat p_s + G (-k_x^2 \hat u_z - k_y^2\hat u_z + \partial_{zz}\hat u_z), \nonumber  \\
\label{hateq}
0 = -ik_x\hat u_x -ik_y\hat u_y + \partial_z \hat u_z,
\end{eqnarray}
subject to the boundary conditions
\begin{eqnarray}
\hat\sigma_{xz}=0=\partial_z \hat u_x -ik_x\hat u_z ~~\hbox{at}~~z=0, \nonumber \\
\hat\sigma_{yz}=0=\partial_z \hat u_y -ik_y\hat u_z ~~\hbox{at}~~z=0, \nonumber \\
\hat\sigma_{zz}=-f=-\hat p_s + 2 G\partial_z\hat u_z ~~\hbox{at}~~z=0, \nonumber  \\
\label{hatbc}
\hat u_x = \hat u_y = \hat u_z = 0 ~~\hbox{at}~~z=-H_l.
\end{eqnarray}
Solving (\ref{hateq}) and (\ref{hatbc}) for $\hat u_z(z=0)$ we find
\begin{equation}
\hat u_z(k_x,k_y,0) = f\frac{2 H_l - \frac{\sinh (2 H_l \sqrt{k_x^2+k_y^2})}{\sqrt{k_x^2+k_y^2}}}
{2 G [1 + 2 H_l^2(k_x^2+k_y^2) + \cosh(2H_l\sqrt{k_x^2+k_y^2})}.
\end{equation} 
Since this corresponds to a radially symmetric integral kernel we can take $q=\sqrt{k_x^2+k_y^2}$
and use the Hankel transform, 
$u_z = \frac{1}{2\pi}\int_0^\infty J_0(r\,q) \,q\, \hat u_z\,dq$,  
to find the surface displacement (Gladwell 1980):
\begin{equation}
\label{dimw}
u_z(r)|_{z=0} = \frac{f}{4 \pi G}
\int_0^\infty J_0(r q) \frac{2 H_l q - \sinh(2 H_l q)}{1 + 2 H_l^2 q^2 + \cosh(2 H_l q)}dq 
= \frac{f}{4 \pi G H_l} W(r),
\end{equation}
where $r = \sqrt{x^2 + y^2}$ and $J_0(r\,q)$ is a Bessel function of the first kind.  
To see the form of the surface displacement, we integrate (\ref{dimw}) numerically and 
show the dimensionless function $W(r)$ in figure \ref{icgf}a.  
The surface displacement $H$ for a pressure distribution $p(x)$ is found using the 
Green's function for a point force (\ref{dimw}), so that
\begin{equation}
\label{incompH}
H = -\int_{-\infty}^\infty \frac{p(x^*)}{4\pi G} \left\{ \int_{-\infty}^\infty 
\left[\int_0^\infty J_0(r q) \frac{2H_lq - \sinh 2H_lq}{1+2 H_l^2 q^2 + \cosh 2H_lq} dq \right]
dy^* \right\} dx^*,
\end{equation}
where $r = \sqrt{(x-x^*)^2 + (y-y^*)^2}$.
In terms of dimensionless variables with the following definitions
\begin{eqnarray}
p = p_0p', ~ q = \frac{q'}{\sqrt{2h_0R}}, ~ x = \sqrt{2h_0R}x', ~ x^* =  \sqrt{2h_0R}x'^*, \nonumber \\
y =   \sqrt{2h_0R}y',~y^* =   \sqrt{2h_0R}y'^*,~ r =    \sqrt{2h_0R} r',~~ h = h_0h', \nonumber \\
H = \frac{p_0 \sqrt{2h_0R}}{4\pi G}H' = \frac{\mu (V-\omega R)}{2\pi G}\frac{R}{h_0}H',
\end{eqnarray}
equation (\ref{incompH}) may be rewritten (after dropping primes) as 
\begin{equation}
\label{icH}
H = -\int_{-\infty}^\infty p(x^*) \left\{ \int_{-\infty}^\infty 
\left[\int_0^\infty J_0(r q) \frac{2\zeta q - \sinh 2\zeta q}{1+2 \zeta^2 q^2 + \cosh 2\zeta q} dq \right]
dy^* \right\} dx^*,
\end{equation}
where $r = \sqrt{(x-x^*)^2 + (y-y^*)^2}$, and the dimensionless group $\zeta = \frac{H_l}{\sqrt{2h_0R}}$
is the ratio of the layer thickness to the size of the contact zone.  To understand the form of (\ref{icH})
we consider the response of a line force $p(x^*) = \delta(x^*)$ and show the results graphically in 
figure \ref{icgf}b.
Moving on to the case at hand, that of a parabolic contact of a cylinder of length $2l$, we write
the dimensionless gap thickness as
\begin{equation}
\label{icg}
h =  1 + x^2 + \eta \, H, ~~ \hbox{where} ~~ \eta =  \frac{1}{2\sqrt2 \pi}\frac{p_0}{G}\sqrt{\frac{R}{h_0}} 
= \frac{1}{2\pi}\frac{\mu (V-\omega R)}{G}\frac{R}{h_0^2} 
\end{equation}
For $\eta\ll1$ we can expand the pressure $p$ in powers of $\eta$ and carry out a 
perturbation analysis as in \S \ref{thine}.
This yields a linear 
relation between the dimensionless lift, $L$, and the scale of the deformation: 
$L = C_i(\zeta) \eta$,
where $C_i(\zeta)$ is shown
in figure \ref{icf}d.
The dimensional lift per unit length is
\begin{equation}
L = \frac{C_i(\zeta)}{2\pi} \frac{\mu^2 (V-\omega R)^2}{G}\frac{R^2}{h_0^3}.
\end{equation}
As $\zeta \to 0$, we approach the limit of an infinitely thick layer, in which case there is no
stiffening due to incompressibility, so that $C(\zeta) \to \frac{3\pi^2}{8}$, which is the result for an 
infinitely thick layer.  To study the effects of confinement on a thick layer, {\it i.e.} $\zeta \gg 1$, 
we approximate (\ref{icH}) as 
\begin{eqnarray}
\label{iH1}
H \approx -\int_{-\infty}^\infty p(x^*) \int_{-l'}^{l'} 
\left\{ \int_0^\infty J_0(r q) [1 -(2+4\zeta q + \zeta^2q^2)e^{-2\zeta q}] dq \right\}
dy^* dx^*,
\end{eqnarray}
where we are now integrating $y^*$ over the dimensionless length of the cylinder $l' = \frac{l}{\sqrt{2h_0R}} \gg 1$,
since the interactions for deep layers are not limited by confinement effects. 
Evaluating (\ref{iH1}) yields
\begin{equation}
\label{iH2}
H \approx \int_{-\infty}^\infty p(x^*) \left\{ \int_{-l'}^{l'} 
\left[ -\frac{1}{r} - \frac{2r^4 + 20r^2\zeta^2+96\zeta^4}{(r^2+4\zeta^2)^{5/2}} \right]
dy^* \right\} dx^*.
\end{equation}
Integrating (\ref{iH2}) with respect to $y^*$ and keeping the leading order terms in $l'$ yields
\begin{eqnarray}
\label{iH3}
H \approx \int_{-\infty}^\infty \frac{2x^*}{(1+x^{*2})^2} \left\{  
\log\left[\frac{4(l^{'2}-y^2)}{(x-x^*)^2}\right] + 2\log\left[\frac{(x-x^*)^2 + 4\zeta^2}{4(l^{'2}+y^2)}\right] \right. \nonumber \\
\left. - \frac{8\zeta^2[(x-x^*)^2+12\zeta^2]}{[(x-x^*)^2+4\zeta^2]^2}
\right\} dx^*,
\end{eqnarray}
where we have used the leading order pressure (\ref{p00}) to evaluate 
$p(x^*) = \frac{2x^*}{(1+x^{*2})^2} + \eta p^{(1)}$.  We integrate (\ref{iH3}) with respect to $x^*$ to find
\begin{equation}
\label{iH4}
H \approx \frac{24\pi x(x^2-1)}{(1+x^2)^2} 
- \frac{4\pi x[x^4+2x^2(1+6\zeta+6\zeta^2)+(1+2\zeta)^2(1+8\zeta+24\zeta^2)]}{[x^2+(1+2\zeta)^2]^3}
\end{equation}
As in \S\ref{thine}, equations (\ref{eta11}) and (\ref{pexpbc}) yields the system of equations to be solved
for the pressure perturbation $p^{(1)}$:
\begin{eqnarray}
\partial_x \left[ 6 H + 3(1+x^2)^2H + (1+x^2)^3\partial_xp^{(1)}\right] = 0 \nonumber \\
\label{iH5}
p^{(1)}(-\infty) = p^{(1)}(\infty) = 0.
\end{eqnarray}
Solving (\ref{iH4})-(\ref{iH5}) yields 
\begin{equation}
p^{(1)}=\frac{2\pi(2x^2-1)}{(1+x^2)^4} - \frac{\pi[12 \zeta^2 - 20\zeta + 30  - x^2(45-60\zeta+36\zeta^2) + 45x^4]}
{2 \zeta^4(1+x^2)^3},
\end{equation}
which we integrate with respect to $x$ to find the dimensionless lift
\begin{equation}
L = \eta p^{(1)} dx = \eta C_i = \eta\frac{3\pi^2}{8}\left(1-\frac{30}{\zeta^4}\right) + O(\zeta^{-5})
\end{equation}

On the other hand, as $\zeta \to 0$ we approach the limit where the contact length is much larger
than the layer thickness leading to geometric stiffening.  
Due to solid incompressibility the layer may only deform via shear and the dominant term of the strain is
$\nabla {\bf u} \sim \frac{u_x}{H_l}$.  Since $\nabla \cdot {\bf u} = 0$ we see that 
$\frac{u_x}{l_c} \sim \frac{H}{H_l} \to \frac{u_x}{H_l} \sim \frac{l_cH}{H_l^2}$.
Balancing the strain energy $\int G (\nabla {\bf u})^2 \, dV \sim G \left( \frac{l_c H}{H_l} \right) H_ll_c$
with the work done by the pressure $p_0Hl_c$ yields $H \sim \frac{p_0H_l^3}{Gl_c^2}$ so that 
$\eta_{\zeta\to \infty} \sim  \frac{p_0H_l^3}{Gl_c^2h_0}$.  For a thin incompressible layer the 
characteristic deflection is reduced by an amount $\frac{\eta_{\zeta \to \infty}}{\eta} \sim \zeta^{-3}$ so 
that $\lim_{\zeta\to\infty} C_i(\zeta) \sim \zeta^{3}$.  
$C_i$ is displayed in figure \ref{icf}a.  
For intermediate values of $\zeta$, we computed the results numerically.

The nonlinear problem arising for $\eta = O(1)$ is solved using (\ref{Req}), (\ref{Reqbc}), (\ref{icH}) and (\ref{icg}). We first guess an initial gap profile $h_{old}$,  then a shooting algorithm is employed to calculate the pressure distribution. The new pressure distribution is then used in (\ref{icH}) and (\ref{icg}) 
to calculate a new gap profile, $ h_{new}$. If $\int_{-10}^{10} (h_{old} - h_{new})^2 dx < 10^{-5}$ 
then the calculation is stopped, else we iterate with $h_{old}=h_{new}$.  
For $\zeta=1$ the results are shown in figure \ref{icf}b-d. Not surprisingly, we find the
same qualitative features discussed previously: $L$ has a maximum due to the competing effects of an increase in the gap thickness and the increased asymmetry of the contact zone.  

\section{Poroelastic layer}
\label{5}

Motivated in part by applications to the mechanics of cartilagenous joints,  we now turn to the case of a cylinder moving above a fluid filled gel layer. This entails a different model for the constitutive behavior of the gel accounting for both the deformation of an elastic network and the fluid flowing through it.  
To describe the mechanical properties of a fluid filled gel we use poroelasticity, the 
continuum description of a material composed of an elastic solid  skeleton and an interstitial fluid  (Biot (1941); for a review of the literature see  Cederbaum, Li \& Schulgasser (2000) or Wang (2000)).  
Our choice of poroelasticity to model the gel is motivated by the following scaling argument 
(Skotheim \& Mahadevan 2004a). Let $\nabla$ and $\nabla_{l_p}$ denote gradients on the system scale and the pore scale respectively; $p_g$ is the pressure varying on the system scale due to boundary conditions driving the flow,  while $p_p$ is the pressure varying on the microscopic scale due to pore geometry. Fluid stress balance on the pore scale implies that the sum of the macroscopic pressure gradient driving the flow, $\nabla p_g$, and the microscopic pressure gradient, $\nabla_{l_p}p_p$, is balanced by the viscous resistance of the fluid having viscosity $\mu$ and velocity ${\bf v}$,  $\mu \nabla_{l_p}^2{\bf v}$, so that the  momentum balance in the fluid yields
\begin{equation}
\label{balape}
\mu \nabla_{l_p}^2{\bf v} - \nabla p_g -\nabla_{l_p}p_p = 0.
\end{equation}
When the pore scale, $l_p$, and system size, $H_l$, are well separated, {\it i.e.} $l_p/H_l \ll 1$, 
equation (\ref{balape}) yields the following scaling relations
\begin{equation}
\label{peconclude}
p_g \sim \frac{H_l \mu V}{l_p^2} \gg \frac{\mu V}{l_p} \sim p_p,
\end{equation}
from which we conclude that the dominant contribution to the fluid stress tensor comes from the pressure. The simplest stress-strain law for the composite medium, proposed by Biot (1941), is found  
by considering the linear superposition of the dominant components of the fluid and solid stress tensor. 
If strains are small, the elastic behaviour of the solid skeleton is well characterized by isotropic
Hookean elasticity.  For a poroelastic material composed of a solid skeleton with Lam\'e coefficients
$G$ and $\lambda$ when drained and a fluid volume fraction $\alpha$,
the stress tensor $\boldsymbol{\sigma}$ is given by the constitutive equation
\begin{equation}
\label{peCE}
\boldsymbol{\sigma} = 
G(\nabla{\bf u} + \nabla{\bf u}^{\hbox{T}}) + \lambda \nabla \cdot {\bf u}\,{\bf \hbox{I}} 
- \alpha p_g{\bf \hbox{I}}.
\end{equation}
The equations of equilibrium are
\begin{equation}
\label{peF}
 \nabla \cdot \boldsymbol{\sigma} =0,
\end{equation}
where we have neglected inertial effects. Mass conservation requires that the rate of dilatation of a solid skeleton having a bulk modulus $\beta^{-1}$ is balanced by the fluid entering the material element:
\begin{equation}
\label{peCont}
\frac{k}{\mu} \nabla^2 p_g  = \beta \partial_t p_g + \partial_t\nabla\cdot{\bf u}.
\end{equation}
$\beta \ne \lambda +2G/3$ since the Lam\'e coefficients $\lambda$ and $G$ are for the composite
material and take into account the microstructure, while $\beta^{-1}$ is independent of the 
microstructure; for cartilage $\beta^{-1} \sim 1$GPa, while $G\sim\lambda\sim1$MPa.
Equations (\ref{peCE}), (\ref{peF}) and (\ref{peCont}) 
subject to appropriate boundary conditions describe the evolution of displacements $\bf u$ and fluid pressure $p_g$ in a poroelastic medium. 

We now calculate the response of a poroelastic 
gel to an arbitrary time dependent pressure distribution before
considering the specific case at hand.
To make the equations dimensionless we use the following scalings 
\begin{eqnarray}
x=\sqrt{2h_0R}x',~~~z=H_lz',
~~~t = \tau t' = \frac{\sqrt{2h_0R}}{V-\omega R} t',
~~~u_z=h_0 u_z', ~~~ 
u_x=h_0 u_x'
\nonumber \\
p_g=p_0p_g'=\mu (V-\omega R)\sqrt{\frac{2R}{h_0^3}}p_g',~~~ p = p_0p',~~~
\boldsymbol{\sigma}=p_0 \boldsymbol{\sigma}' = \mu (V-\omega R) 
\sqrt{\frac{2R}{h_0^3}}\boldsymbol{\sigma}'.
\end{eqnarray}
We take the thickness of the layer to be much smaller that the length scale of the contact zone, 
$\zeta = \frac{H_l}{\sqrt{2 h_0R}} \ll 1$, and consider a compressible material, $G \sim \lambda$.
Then after dropping primes, the stress tensor (\ref{peCE}) becomes
\begin{equation}
\boldsymbol{\sigma}=
\left( 
\begin{array}{cc}
\frac{1}{\eta} \frac{\lambda}{2G+\lambda} \partial_z u_z + \frac{\zeta}{\eta}\partial_x u_x -\alpha p_g &
\frac{1}{\eta}\frac{G}{2G+\lambda} \partial_z u_x + \frac{\zeta}{\eta}\frac{G}{2G+\lambda} \partial_x u_z \\
\frac{1}{\eta}\frac{G}{2G+\lambda} \partial_z u_x + \frac{\zeta}{\eta}\frac{G}{2G+\lambda} \partial_x u_z & 
\frac{1}{\eta} \partial_z u_z + \frac{\zeta}{\eta} \frac{\lambda}{2G+\lambda} \partial_x u_x - \alpha p_g
\end{array}
\right)
\end{equation}
Here,  
\begin{equation}
\label{peeta}
\eta =\frac{p_0}{2G+\lambda} \frac{H_l}{h_0} =
\sqrt{2} \frac{\mu (V-\omega R)}{2G + \lambda} \frac{ H_l R^{1/2}}{h_0^{5/2}},
\end{equation}
is the dimensionless number 
governing the relative size of the surface deflection to the undeformed gap thickness,
{\it i.e.} the material compliance.  
Stress balance (\ref{peF}) yields 
\begin{eqnarray}
0=\partial_{zz}u_x + \zeta \left( \frac{G+\lambda}{G}\partial_{xz}u_z 
-\eta \alpha \frac{2G + \lambda}{G} \partial_xp_g \right) + \zeta^2\frac{2G + \lambda}{G}\partial_{xx}u_x,
\nonumber \\
\label{pefb}
0=\partial_{zz}u_z - \eta \alpha \partial_z p_g + \zeta \frac{G+\lambda}{2G + \lambda} \partial_{xz}u_x
+\zeta^2 \frac{G}{2G+\lambda}\partial_{xx}u_z,
\end{eqnarray}
and continuity (\ref{peCont}) yields
\begin{equation}
\label{pec2}
\frac{k \tau}{\mu H_l^2 \beta}(\partial_{zz}p_g + \zeta^2\partial_{xx}p_g)
= \partial_t p_g + \frac{h_0}{\beta H_l p_0} \partial_t(\partial_z u_z + \zeta\partial_x u_x).
\end{equation}
To leading order (\ref{pefb}) and (\ref{pec2}) reduce to 
\begin{eqnarray}
\partial_{zz} u_x = 0,
\nonumber \\
\label{pevert}
\partial_{zz} u_z -\alpha \eta \partial_z p_g=0,
\end{eqnarray}
and 
\begin{equation}
\label{pec3}
\frac{k \tau}{\mu H_l^2 \beta}\partial_{zz}p_g
= \partial_t p_g + \frac{h_0}{\beta H_l p_0} \partial_{tz} u_z.
\end{equation}
The normal unit vector to the soft interface 
is ${\bf n} = (-\partial_x u_z|_{z=0}\,,1)$, which in 
dimensionless form is 
\begin{equation}
{\bf n} = ( -\varepsilon \partial_x u_z|_{z=0},1),
\end{equation}
where $\varepsilon = \sqrt\frac{h_0}{2R}$.
The balance of normal traction on the solid-fluid interface yields 
\begin{equation}
\label{surface}
\boldsymbol{\sigma}_f\cdot{\bf n}|_{z=0} = \boldsymbol{\sigma}_s\cdot{\bf n}|_{z=0},
\end{equation}
so that 
\begin{eqnarray}
\partial_z u_x|_{z=0} =0 \nonumber \\
\label{peuzsurf}
  \partial_z u_z|_{z=0} - \alpha \eta p_g|_{z=0}= - \eta \, p.
\end{eqnarray}
At the interface between the soft film and the underlying rigid substrate the no slip condition yields
\begin{equation}
\label{pevertbc}
u_z(x,-1)=0, ~~~ u_x(x,-1)=0.
\end{equation}
Solving (\ref{pevert}), (\ref{peuzsurf}) and (\ref{pevertbc}) yields
\begin{eqnarray}
u_x = 0, \nonumber \\
\label{peu_z}
-\eta \, p=-\alpha \, \eta \, p_g+\partial_z u_z.
\end{eqnarray}
To calculate the displacement at the surface $u_z(x,0,t)$ we need to determine the 
fluid pressure in the gel $p_g$.
Using (\ref{peu_z}) in (\ref{pec3}) yields
\begin{equation}
\label{17}
\partial_tp_g-\gamma\partial_{zz}p_g = \delta \partial_tp
\end{equation}
where
\begin{equation}
\label{pegamma}
\gamma = \frac{\tau}{\tau_p} = \frac{k\sqrt{2Rh_0}(2G+\lambda)}{H_l^2(V-\omega R)[\beta(2G+\lambda)+\alpha]}
~~~\hbox{and}~~~\delta=\frac{1}{\beta(2G+\lambda)+\alpha} \sim O(1).
\end{equation}
Here $\tau = \frac{l_c}{V} \sim \frac{\sqrt{h_0R}}{V}$ 
is the time scale associated with motion over the contact length,
$\tau_p \sim \frac{\mu H_l^2}{k\,G}$ is the time scale associated with stress relaxation
via fluid flow across the thickness of the gel,
and $\delta \sim \frac{1}{\alpha}$ the inverse of the fluid volume fraction.
Equation (\ref{17}) corresponds to the short time limit discussed by Barry and Holmes (2001).
The boundary conditions for (\ref{17}) are determined by the fact that 
at the solid-gel interface there is no fluid flux and 
at the fluid-gel boundary there is no pressure jump so that  
\begin{equation}
\label{diff}
\partial_zp_g(x,-1,t) = 0, ~~~ p_g(x,0,t)=p,
\end{equation}
where both conditions are a consequence of Darcy's law for flow through a porous medium.  

Although there is flux through the gel-fluid interface, 
the Reynolds equation (\ref{Req}) for the fluid pressure will remain 
valid if the fluid flux through the gel is much less than the flux through the thin gap.
A fluid of viscosity $\mu$ 
flows with a velocity ${\bf v} = (v_x,v_z)$ through a porous medium of isotropic permeability $k$
according to Darcy's law, so that ${\bf v} \sim \frac{k}{\mu} \nabla p$.
Hence, the total flux through a porous medium of thickness $H_l$ is $\int_{-H_l}^0v_x\,dz$, which
will scale as $\frac{H_l\,k\,p_0}{\mu \sqrt{h_0R}} \sim \frac{H_l\,k\,V}{\mu h_0^2}$.  
Comparing this with the flux through the thin gap 
$h_0 V$ leads to the dimensionless group $Q_R = \frac{H_l\,k}{h_0^3}$. 
If $Q_R\ll1$ we can neglect the flow through the 
porous medium.  For cartilage, $H_l\sim 1mm$ and $k \sim10^{-14}mm^2$, and flow
through the porous medium can be neglected if $h_0\gg10nm$.
This implies that the Reynolds lubrication approximation embodied in (\ref{Req}) remains valid in 
the gap for situations of biological interest.

In response to forcing, a poroelastic material can behave in three different ways depending
on the relative magnitude of the time scale of the motion $\tau = l_c/V$ 
and the poroelastic 
time scale $\tau_p$.
If $\tau\gg\tau_p$ the fluid in the gel is always in equilibrium with the surrounding fluid and 
a purely elastic theory for the deformation of the gel suffices; 
if $\tau\sim\tau_p$ the gel will behave as a material with a 'memory'; 
if $\tau\ll\tau_p$ the fluid has no time to move relative to the
matrix and the poroelastic material will again behave as a solid albeit with a higher elastic modulus.
In the physiological case of a cartilage layer 
in a rotational joint the poroelastic time scale for bovine articular cartilage is reported to be
$\tau_p \approx 20$ seconds by Grodzinsky, Lipshitz \& Glimcher (1978), and 
$\tau_p \approx 500$ seconds by Mow, Holmes \& Lai (1984). 
For time scales on the order of 1 second, 
the cartilage should behave as a solid, 
but with an elastic modulus 
greater then that measured by equilibrium studies.  
We consider three different cases corresponding to:

\subsection{Low speed:  $\tau_m \gg \tau_p$}

When the cylinder moves slowly, $\tau_m \gg \tau_p$, 
the time scale of the motion is much larger than the time scale
over which the pressure diffuses 
({\it i.e.} $\gamma \gg 1\sim \delta$) so that (\ref{17}) becomes
\begin{equation}
\label{7.9}
\partial_{zz}p_g=0
\end{equation}
Solving (\ref{7.9}) subject to (\ref{diff}) yields
\begin{equation}
p_g=p,
\end{equation}
{\it i.e.} at low speeds the fluid pressure in the gel is the same as the fluid pressure outside the gel.
Equations (\ref{pevertbc}) and (\ref{peu_z}) can be solved to yield
\begin{equation}
u_z(x,0,t) = -(1-\alpha) \eta p(x,t).
\end{equation}
We see that this limit gives a local relationship between the displacement of the gel surface 
and the fluid pressure in the gap, exactly as in the case of a purely elastic 
layer treated in \S \ref{thine}.

\subsection{High speed: $\tau_m \ll \tau_p$ }

When the time scale of the motion is much smaller than
the time scale over which the pressure diffuses, {\it i.e.} $\gamma \ll 1\sim \delta$, (\ref{17}) becomes
\begin{equation}
\label{fast}
\partial_tp_g = \delta\partial_tp.
\end{equation}
Since the gel is at equilibrium with the external fluid before the cylinder passes over it,
$p_g(x,z,-\infty)=p(x,-\infty)=0$, and equation (\ref{fast}) yields
\begin{equation}
\label{pg}
p_g(x,z,t)=\delta\,p(x,t)
\end{equation}
Inserting (\ref{pg}) into (\ref{peu_z}) and integrating yields
\begin{equation}
u_z(x,0,t) = \frac{-\beta(2G+\lambda)\eta}{\beta(2G+\lambda)+\alpha}\,p(x,1,t).
\end{equation}
In this limit the fluid has no time to flow through the pores and the only compression is due to bulk
compressibility of the composite gel, 
which now behaves much more rigidly.  
The effective elastic modulus of the solid layer is 
now $G_{eff} \sim \beta^{-1} \sim 1$GPa rather than $G_{eff} \sim 1$MPa.
However, the relationship between the pressure and
displacement remains local as in \S \ref{thine}.
  
We note that if $\nabla\cdot {\bf u}=0$, (\ref{peCont}) has no forcing term and $p_g=0$.  
Poroelastic theory does not take into account shear deformations
since these involve no local change in fluid volume fraction in the gel.  
In this case all the load
will be borne by the elastic skeleton.  However, shear deformation in a thin layer will involve 
geometric stiffening due to incompressibility so that the effective modulus will be 
$G_{eff} \sim \frac{h_0R}{H_l^2}G$ (Skotheim \& Mahadevan 2004b).  Hence, if 
$\frac{h_0R}{H_l^2}G \beta \ll 1$ the deformation should be treated as an incompressible 
layer as in \S\ref{4}.  If $\frac{h_0R}{H_l^2}G \beta \gg 1$, the layer should be treated as in \S\ref{thine}
with an effective modulus $\beta^{-1}$.

\subsection{Intermediate speeds: $\tau_m \sim \tau_p$}

\begin{figure}
\begin{center}
\includegraphics[width=6cm]{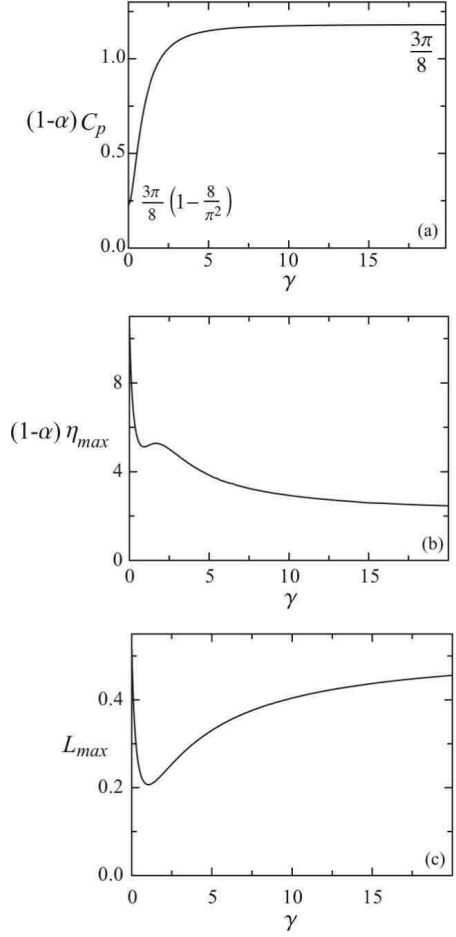}
\par
\caption{For $\eta \ll 1$, $L=C_p(\gamma)\, \eta$
where $\gamma$ is the ratio of translational to poroelastic time scales, and 
$C_p(\gamma)$ is shown in (a).
(b) $\eta_{max}$, the value of $\eta$ at which the lift is maximum, plotted against $\gamma$.
(c) The maximum lift $L_{max}$ 
as a function of $\gamma$.  
After scaling $L$ and $\eta$, $\frac{L[\frac{\eta}{\eta_{max}(\gamma)}]}{L_{max}(\gamma)}$, the 
curves can be almost perfectly collapsed onto a single curve.}
\label{maxporo}
\end{center}
\end{figure}

When $\delta\sim\gamma\sim 1$ rewriting (\ref{17}) for the
difference between the fluid pressure inside and outside the gel, $p_g(x,z,t) - p(x,t)$, yields
\begin{equation}
\label{above}
\partial_t(p_g-p) - \gamma\partial_{zz}(p_g-p) = (\delta-1)\partial_tp,
\end{equation}
with the boundary conditions
\begin{equation}
\label{above2}
\partial_z (p_g-p)(x,-1,t) = 0, ~~~ (p_g - p)(x,0,t)=0.
\end{equation}
We expand $p_g-p$ in terms of the solution of the homogeneous part of (\ref{above})-(\ref{above2}): 
\begin{equation}
\label{homog}
p_g - p= \sum_{n=0}^\infty A_n(t) \sin \pi(n+\frac{1}{2})z.
\end{equation}
Inserting the expansion into (\ref{above}) we find
\begin{equation}
\label{7.18}
\sum_{n=0}[\partial_t A_n + \gamma \pi^2(n+\frac{1}{2})^2A_n]\sin \pi(n+\frac{1}{2})z = (\delta-1)\partial_tp.
\end{equation}
Multiplying (\ref{7.18})  with $\sin\pi(m+\frac{1}{2})z$ and integrating over the thickness yields
\begin{equation}
\label{7.19}
\partial_t A_n + \gamma \pi^2(n+\frac{1}{2})^2A_n = \frac{2(1-\delta) }{\pi(n+\frac{1}{2})}
\partial_t p.
\end{equation}
Solving (\ref{7.19}) for $A_n(t)$ yields 
\begin{equation}
\label{7.20}
A_n(t) = \frac{2(1-\delta)}{\pi(n+\frac{1}{2})}\int_{-\infty}^te^{-\gamma\pi^2(n+\frac{1}{2})^2(t-t')}\partial_{t'}p\,dt'.
\end{equation}
Substituting (\ref{7.20}) into (\ref{homog}) yields the fluid pressure in the gel 
\begin{equation}
\label{pgel}
p_g=p+\sum_{n=0}\frac{2(1-\delta)\sin\pi(n+\frac{1}{2})z}{\pi(n+\frac{1}{2})}
\int_{-\infty}^te^{-\gamma\pi^2(n+\frac{1}{2})^2(t-t')}\partial_{t'}p\,dt'.
\end{equation}
Finally, (\ref{pevertbc}), (\ref{peu_z}) and (\ref{pgel}) yield
\begin{equation}
\label{25}
u_z(1) = \eta\left[ (-1+\alpha)p + \alpha\sum_{n=0}\frac{2(\delta-1)}{\pi^2(n+\frac{1}{2})^2}\int_{-\infty}^te^{-\gamma\pi^2(n+\frac{1}{2})^2(t-t')}\partial_{t'}p\,dt'\right].
\end{equation}
Since the higher order diffusive modes ($n>0$) decay more rapidly than the leading order diffusive 
mode ($n=0$), a good approximation to (\ref{25}) is
\begin{equation}
\label{7.23}
-\eta H(x) = u_z(0) = \eta\left[(-1+\alpha)p + \alpha\frac{8(\delta-1)}{\pi^2}\int_{-\infty}^te^{\frac{-\gamma\pi^2}{4}(t-t')}\partial_{t'}p\,dt'\right].
\end{equation}
This approximation is similar to that used in Skotheim \& Mahadevan (2004a).
To simplify (\ref{7.23}) for the case of interest we define
\begin{equation}
\label{pe102}
\xi = \int_{-\infty}^t e^{\frac{-\gamma\pi^2}{4}(t-t')}\partial_{t'}p\,dt'.
\end{equation}
so that
\begin{equation}
\partial_t \xi = -\frac{\gamma\pi^2}{4}\xi + \partial_t p.
\end{equation}
In the reference frame of the steadily moving cylinder $\xi(x,t)=\xi(x-t)$ so that
\begin{equation}
\partial_x \xi = \frac{\gamma\pi^2}{4}\xi + \partial_x p.
\end{equation}
Integrating the above yields
\begin{equation}
\label{pe105}
\xi = -\int_x^\infty e^{ \frac{-\gamma\pi^2}{4}(x'-x)}\partial_{x'}p\,dx'
\end{equation}
Consequently, 
the distance between the gel and cylinder is found from (\ref{h(x)}), (\ref{7.23}), (\ref{pe102}) and 
(\ref{pe105}) to be
\begin{equation}
\label{h1}
h(x)=1+x^2+ \eta (1-\alpha)[p + \frac{8}{\pi^2} \int_{x}^\infty e^{\frac{-\gamma \pi^2}{4} (x'-x)}\partial_{x'}p\,dx']
\end{equation}
where 
\begin{equation}
\label{pe3d}
\eta = \frac{p_0 H_l}{(2G+\lambda) h_0} = \frac{\sqrt{2R} H_l\mu (V-\omega R)}{h_0^{5/2}(2G+\lambda)},
\end{equation}
where we are considering the case where 
the bulk modulus of the skeletal material is much larger than the modulus of the
elastic matrix $\beta G\ll1$, so that (\ref{pegamma}) implies $\delta \approx \frac{1}{\alpha}$.  
This leaves us a system of equations (\ref{Req}), (\ref{Reqbc}) and (\ref{h1}) 
for the pressure with 2 parameters: $\eta$ characterizes the deformation (softness); 
and $\gamma$ is the ratio of translational to diffusive timescales.
The two limits $\gamma \ll 1$ and $\gamma \gg 1$ of (\ref{h1}) can both be treated using 
asymptotic methods.   
For $\gamma \gg 1$, (\ref{h1}) yields
\begin{equation}
h = 1+x^2+\eta (1-\alpha) \,p,
\end{equation}
and we recover the limit of a thin compressible elastic layer treated in \S\ref{thine} with  
$\eta\to (1-\alpha) \eta$.  
For $\gamma \ll 1$, (\ref{h1}) yields
\begin{equation}
h = 1 + x^2 + (1-\frac{8}{\pi^2})(1-\alpha)\eta\,p,
\end{equation}
which is the result for a thin compressible layer with $\eta\to(1-\frac{8}{\pi^2})(1-\alpha)\eta$.
When $\eta \ll \gamma \ll 1$ we expand the pressure field
as in (\ref{expansion}) writing
$p = p_0 + \eta p_1 $, 
where $p_0 = \frac{2x}{(1+x^2)^2}$ as in (\ref{p00}).    Inserting this expression
into (\ref{h1}) yields
\begin{equation}
h = 1+x^2 + \eta (1-\alpha) \left\{ p_0 + \frac{8}{\pi^2}\int_x^\infty [1 + \frac{\pi^2 \gamma}{4} (x-x')] \partial_{x'}p_0\,dx' \right\} + O(\gamma \eta)
\end{equation}
which can be integrated to give
\begin{equation}
h = 1 + x^2 + \eta (1-\alpha) \left[ \frac{(\pi^2-8)2x}{\pi^2(1+x^2)^2} 
+ \frac{2 \gamma}{(1+x^2)}\right].
\end{equation}
We see that increasing $\gamma$ increases the gap thickness and lowers the
pressure without increasing
the asymmetry, thus decreasing the lift.
In the small deflection limit, $\eta \ll 1$, the dimensionless lift force is $L = C_p(\gamma) \eta$ 
and the lift force in dimensional terms is
\begin{equation}
L = C_p(\gamma)(1-\alpha)\frac{\mu^2V^2}{2G+\lambda}\frac{H_lR^{3/2}}{h_0^{7/2}},
\end{equation}
where $C_p$ is a function of $\gamma$ and shown graphically in figure \ref{maxporo}a.

When $\eta = O(1)$, we use a numerical method to 
solve (\ref{Req}), (\ref{Reqbc}) and (\ref{h1}) on a finite domain 
using the continuation software AUTO (Doedel {\it et al.} 2004) with 
$\eta$ and $\gamma$ as the continuation parameters.  
The initial solution 
from which the continuation begins is with $\eta=\gamma=0$, corresponding to 
%
$h = 1+ x^2$,  and  $p = \frac{2x}{(1+x^2)^2}$.
%
The form of the lift force as a function of $\eta$, $L(\eta,\gamma)$ for various $\gamma$ can be 
almost perfectly collapsed onto a single curve after appropriately scaling the $\eta, L$ axes
using the position of the maximum;
{\it i.e.} $\frac{L[\eta/\eta_{max}(\gamma)]}{L_{max}(\gamma)}$ where $L_{max}$ and $\eta_{max}$
are shown in figure \ref{maxporo}b,c.

\section{Elastic shell}
\label{7}

\begin{figure}
\begin{center}
\includegraphics[width=10cm]{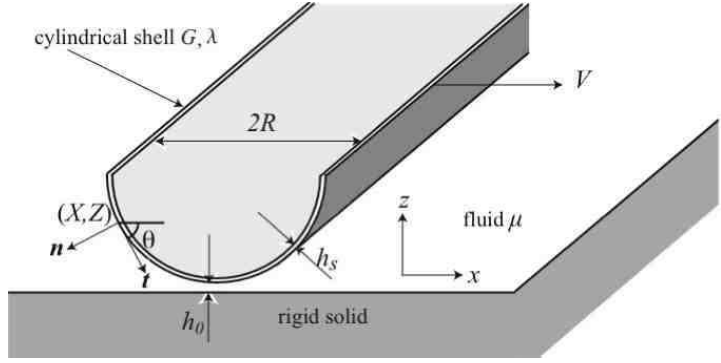}
\par
\caption{Schematic diagram of a half-cylinder of radius $R$, thickness $h_s$ and Lam\'e 
coefficients $\mu$ and $\lambda$ moving at a velocity $V$ while completely immersed
in a fluid of viscosity $\mu$.   
The edges of the half-cylinder are clamped at a distance $R + h_0$ from the surface of 
an undeformed solid. $\theta$ denotes the angle between the tangent to the surface and the 
$x-$axis.  $(X(s),Z(s))$ are the laboratory frame coordinates of the half-cylinder as a function of 
the arc-length coordinate $s$.}
\label{esschematic}
\end{center}
\end{figure}

\begin{figure}
\begin{center}
\includegraphics[width=10cm]{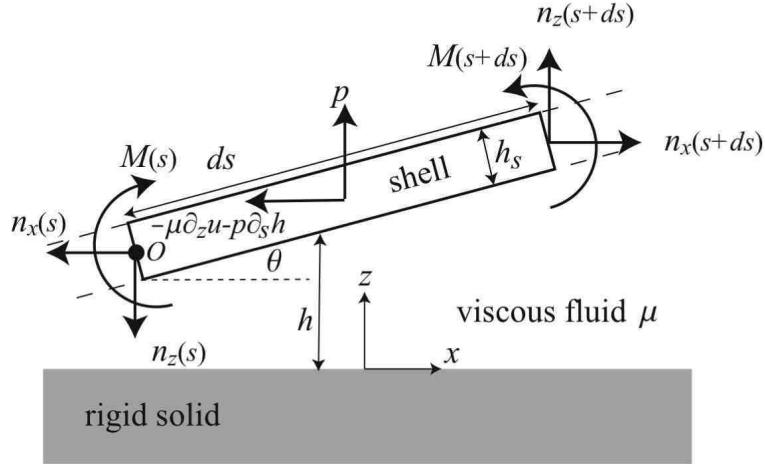}
\par
\caption{Schematic of the torque and force balance for a bent cylindrical shell of thickness
$h_s$ and Lam\'e coefficients $G$ and $\lambda$ subject to a traction $(-\mu\partial_zu-p\partial_xh,p)$
applied by a viscous fluid.
$x$ and $z$ are coordinates in the reference frame
of the rigid solid, while $s$ is the arc-length coordinate in the shell.  
$M(s)=\frac{G(\lambda+G)h_s^3}{3(\lambda+2G)}\partial_{ss}\theta$ 
is the bending moment.}
\label{elastica}
\end{center}
\end{figure}

For elastic layers attached to a rigid substrate the effective stiffness increases with decreasing thickness.
However, for free elastic shells the effective stiffness increases with increasing thickness.  
To see the effects of this type of geometry in biolubrication problems,
we turn our attention to a configuration in which a surface is rendered soft through
its geometry rather than its elastic moduli. 
We consider a half-cylindrical elastic shell of thickness $h_s$ and radius $R$
moving with constant 
velocity $V$ parallel to the rigid substrate, while completely immersed in fluid of viscosity $\mu$, 
as shown in figure \ref{esschematic}.  
The shell is clamped at its edges, which are at a height $R + h_0$ above the rigid solid.
The shape of the elastic half-cylinder is governed by the {\it elastica} equation 
(for the history of the {\it elastica} equation as well as its derivation see Love 1944) 
for $\theta(s)$, the angle between the horizontal and the tangent vector where $s$ is the arc length coordinate.
Balancing torques about the point $O$ in figure \ref{elastica} gives: 
$M(s+ds)-M(s)-ds\,n_x\sin\theta +ds\,n_z\cos\theta + \frac{ds^2}{2}
(p\cos\theta -  \mu\partial_z u\sin\theta - p\partial_s h\sin\theta) = 0$.  
In the limit $ds\to0$, $\partial_xM -n_x \sin\theta+n_z\cos\theta=0$. 
The $x-$force balance is $n_x(s+ds)-n_x(s) - ds(p\partial_sh + \mu\partial_xu)=0$, which as 
$ds\to0$ yields $\partial_sn_x = \mu\partial_xu + p\partial_sh$.
The $z-$force balance is $n_z(s+ds)-n_z(s) + p\,ds=0$, which as 
$ds\to0$ yields $\partial_sn_z = -p$.  We note that external forces are applied in the contact region 
where derivatives taken with respect to $x$ are interchangeable with those taken with respect to $s$, 
{\it i.e.} $\partial_x h \approx \partial_s h$ and $\partial_x p \approx \partial_s p$, which allows for 
a consistent framework for the fluid and solid equilibrium equations.
This yields 
\begin{eqnarray}
\label{elastica1}
\frac{G(\lambda+G) h_s^3}{3(\lambda+2\mu)}\partial_{ss}\theta = n_x\sin\theta - n_z \cos\theta, 
\end{eqnarray}
where the stress resultants $n_x,~n_z$ are determined 
by the equations
\begin{eqnarray}
0=\partial_s n_x + {\bf e_x}\cdot \boldsymbol{\sigma}_f \cdot {\bf n},~~~~
0=\partial_s n_z  + {\bf e_z}\cdot \boldsymbol{\sigma}_f  \cdot {\bf n} 
\end{eqnarray}
{\it i.e.}
\begin{eqnarray}
\label{e1}
\partial_s n_x =  \frac{\mu V}{h} + \frac{h\partial_sp}{2} + p\partial_sh,
~~~~\partial_s n_z =  -p, 
\end{eqnarray}
where, ${\bf n} = (\partial_s h,-1)$, and $\boldsymbol{\sigma}_f $ is given by (\ref{fst1}).  
Then, the pressure in the fluid is governed by the Reynolds equation~(\ref{Req})
\begin{equation}
\label{r1}
\partial_{ss}p = \frac{-\partial_sh}{h^3}(6\mu V + 3h^2\partial_s p).
\end{equation}
Finally, since cylindrical deformations are inextensional, we must complement (\ref{elastica1})-(\ref{r1})
with the kinematic equations
\begin{equation}
\label{kin}
\partial_s X = \cos\theta,~~~\partial_s h = \sin\theta.
\end{equation}
where the position of the surface of the elastic cylinder is $(X,h)$. 
The equations (\ref{elastica1})-(\ref{kin}) are made dimensionless with the following scalings
\begin{eqnarray}
p = p_0p' = \frac{\sqrt{2}\mu V R^{1/2}}{h_0^{3/2}} p', ~~~ s = \pi R s', ~~~ h = R\,Z,  ~ \nonumber \\ 
X = R\,X' ~~~~
n_x = p_0h_0 n_x', ~~~~
n_z = p_0\sqrt{2h_0R}\,n_z'.  
\end{eqnarray} 
After dropping primes, the dimensionless forms of equations (\ref{elastica1})-(\ref{kin}) are written as
\begin{eqnarray}
\label{ae1}
\partial_{ss}p& = & \frac{-\partial_sZ}{Z^3}[3\sqrt{2}\pi(\frac{h_0}{R})^{3/2}  + 3Z^2\partial_s p], \\
\partial_s n_x& = & \frac{R}{h_0}(p\partial_sZ + \frac{Z\partial_s p}{2}) + \sqrt{\frac{h_0}{2R}}\frac{\pi}{Z} , \\
\partial_s n_z& = & -\pi\sqrt{\frac{R}{2h_0}}p, \\
\partial_s X& = & \pi\cos\theta, \\
\label{aef}
\partial_s Z& = & \pi\sin\theta, \\
\label{ff}
\partial_{ss} \theta& = & 3\sqrt{2}\pi^2 \frac{\mu V(\lambda+2G)}{G(\lambda+G)} \frac{R^{5/2}}{h_s^3h_o^{1/2}} (n_x\sin\theta - \sqrt{\frac{2 R}{h_0}} n_z\cos\theta)
\end{eqnarray}
To find the scale of the elastic deformation $\eta = \frac{H_0}{h_0}$, where 
the maximum displacement of the cylinder is of order $H_0$, 
we note that  the change in curvature is of order $\frac{H_0}{R^2}$ so that 
the bending strain is $\epsilon = \frac{h_s H_0}{R^2}$ and the 
elastic energy per unit length therefore scales as 
$\int G \epsilon^2 dA \sim \int G (\frac{h_s H_0}{R^2})^2 dA \sim \frac{G\,h_s^3 H_0^2}{R^3}$.
The work done by the fluid is due to a localized torque and scales as 
$\int p\,x\,dx\sim p_0\,h_0R$, acting through an angle $\Delta\theta\sim \frac{H_0}{R}$.  
Balancing the work done by the fluid torque $p_0\,h_0 H_0$ with the elastic energy 
$\frac{G\,h_s^3 H_0^2}{R^3}$ yields
\begin{equation}
\label{escale}
\eta \sim \frac{H_0}{h_0} \sim \frac{\mu V}{G} \frac{R^{7/2}}{h_s^3h_0^{3/2}},
\end{equation}
so that (\ref{ff}) can be written as
\begin{equation}
\label{tf}
\partial_{ss} \theta = \frac{h_0\eta}{R} (n_x\sin\theta - \sqrt{\frac{2 R}{h_0}} n_z\cos\theta).
\end{equation}
The system (\ref{ae1})-(\ref{aef}), (\ref{tf}) for $\theta, p, n_x, n_z, x, z$ has
two dimensionless parameters, 
$\eta  = 3\sqrt{2}\pi^2 \frac{\mu V(\lambda+2G)}{G(\lambda+G)} \frac{R^{7/2}}{h_s^3h_0^{3/2}} $ 
and $\frac{h_0}{R}$,
and is subject to 8 boundary conditions
\begin{eqnarray}
p(0)&=&p(1)=0, \nonumber \\
X(0)&=&-1, ~~~ X(1) = 1, \nonumber \\
Z(0)&=&Z(1) =1 + \frac{h_0}{R}, \nonumber  \\
\label{shellbc}
\theta(0)&=&\frac{-\pi}{2}, ~~~ \theta(1) = \frac{\pi}{2}.
\end{eqnarray}
The first two are a consequence of lubrication theory, while the rest are kinematic boundary conditions
on the cylinder's lateral edges, which are assumed to be clamped. 
We note that if the cylinder has a natural curvature this will not effect the system of equations 
(\ref{elastica1})-(\ref{kin}), but may change the boundary conditions.

\begin{figure}
\begin{center}
\includegraphics[width=10cm]{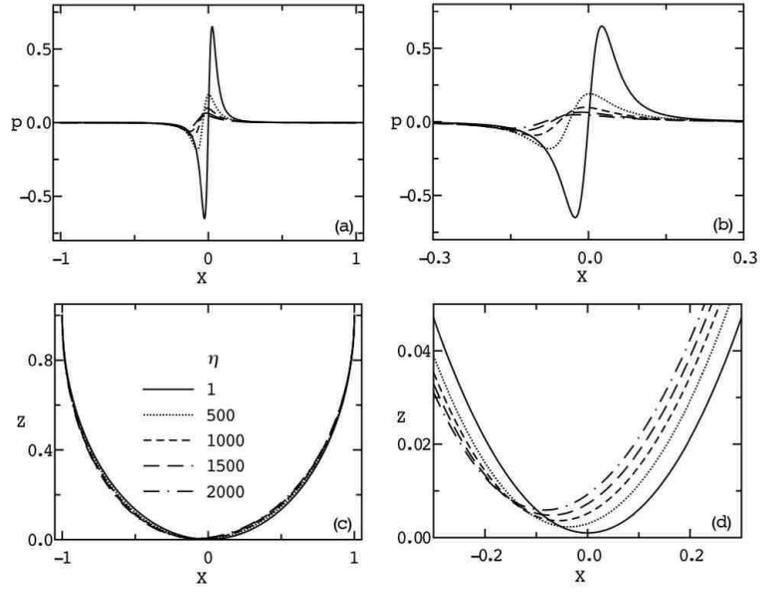}
\par
\caption{(a),(b)
pressure distribution, $p(X)$, as a function of the softness, $\eta$.  
As $\eta$ increases the asymmetry of the
pressure distribution increases and the maximum pressure decreases.  
(c),(d) shape of the sheet, where $X(s)$ and $Z(s)$ are the coordinates of the center line in the laboratory frame.  We see that the point of nearest contact is pulled back and the symmetry of the 
profile is broken by the forces exerted by the fluid on the cylindrical shell. $\frac{h_0}{R}=10^{-3}$.}
\label{cylshape}
\end{center}
\end{figure}

\begin{figure}
\begin{center}
\includegraphics[width=10cm]{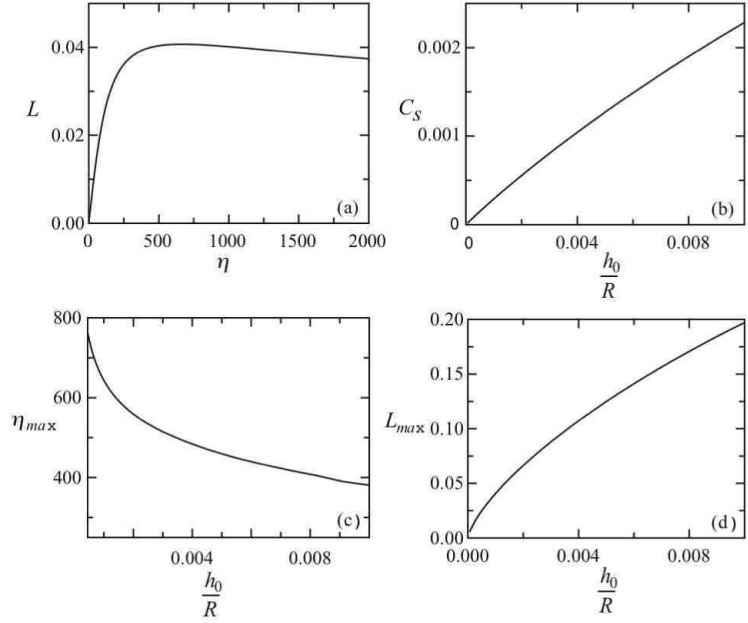}
\par
\caption{(a) dimensionless lift for a cylindrical shell for $\frac{h_0}{R}=10^{-3}$.  For $\eta < 100$.
$L = C_s(\frac{h_0}{R}) \eta$, where $C_s$ is shown in (b).
The form of $L(\eta,\frac{h_0}{R})$ can be almost perfectly collapsed onto a single curve after appropriately scaling 
the $\eta$, $L$ axes, {\it i.e.} $\frac{L[\frac{\eta}{\eta_{max}(h_0/R)}]}{L_{max}(h_0/R)}$, where $\eta_{max}$ and
$L_{max}$ are shown in (c) and (d) respectively.
}
\label{asymptotic}
\end{center}
\end{figure}

We solve (\ref{ae1})-(\ref{aef}),  (\ref{tf}) and (\ref{shellbc}) numerically using 
the continuation software AUTO 2000 (Doedel {\it et al.} 2004) with $\eta$ as the continuation parameter. 
For $\eta < 1$,  the dimensionless lift $L = C_s \eta$,
where the constant $C_s(\frac{h_0}{R})$ is shown in figure \ref{asymptotic}b.  However, we see that even 
for $\eta<100$ the linear relationship remains valid.
The dimensional lift force for $\eta<100$ is 
\begin{equation}
L = 6\sqrt{2}\pi^2\, \frac{\mu^2 V^2(\lambda+2G)}{G(\lambda+G)} \frac{R^{9/2}}{h_s^3h_0^{5/2}}C_s\left(\frac{h_0}{R}\right).
\end{equation}
Figure \ref{cylshape}  shows the gap thickness profile and pressure distribution 
and figure \ref{asymptotic}a shows the dimensionless lift for  
$\frac{h_0}{R}=0.001$.  
As $\eta$ increases the elastic deformation breaks the symmetry of the
gap thickness profile and results in an asymmetric pressure distribution and corresponding lift.  
However, the concomitant increase in gap thickness decreases the magnitude of the pressure.
As for the previous systems considered, the competition between symmetry breaking 
(dominant for small $\eta$) and decreasing pressure (dominant for large $\eta$) produces 
a maximum in the lift. 
The form of the lift force as a function of $\eta$, $L(\eta,\frac{h_0}{R})$ for various $\frac{h_0}{R}$ 
can be almost perfectly collapsed onto a single curve after appropriately scaling the 
$\eta, L$ axes;
{\it i.e.} $\frac{L[\eta/\eta_{max}(\frac{h_0}{R})]}{L_{max}(\frac{h_0}{R})}$ 
where $L_{max}$ and $\eta_{max}$
are shown in figure \ref{asymptotic}c,d.

\section{Journal bearing}
\label{8}

\begin{figure}
\begin{center}
\includegraphics[width=9cm]{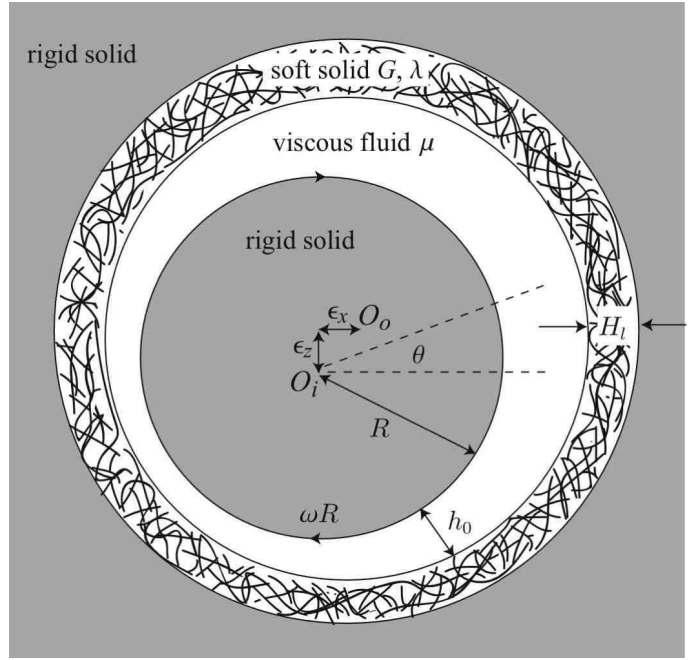}
\par
\caption{Schematic diagram of the modified journal bearing geometry in which the larger cylinder of
radius $R+H_l+h_0$
has been coated by a soft solid of thickness $H_l$ having Lam\'e coefficients $G$ and $\lambda$.
The larger cylinder's axis is located a distance $\epsilon_x$ in the $x-$direction and a distance 
$\epsilon_z$ in the $z-$direction from the axis of the inner cylinder of radius $R$. The average 
gap thickness is $h_0$.
}
\label{jbschematic}
\end{center}
\end{figure}

So far, with the exception of \S\ref{Dsec}, we have dealt only with non-conforming contacts.
In this section we consider an elastohydrodynamic journal bearing: a 
geometry consisting of a cylinder rotating within a larger cylinder that is coated with a soft solid.  
The journal bearing is a conforming contact and 
is a better representation of bio-lubrication in mammalian 
joints in which synovial fluid lubricates bone 
coated with thin soft cartilage layers. 
Previous analyses of the elastohydrodynamic journal bearing have focused on situations where fluid
cavitation needs to be taken into account 
(O'Donoghue, Brighton \& Hooke 1967; Conway \& Lee 1975).  
As before, we restrict our attention the case where the surface deforms appreciably before 
the cavitation threshold is reached so that the gap remains fully flooded.
A schematic diagram is shown in figure \ref{jbschematic}.   

We take the center of the inner cylinder, $O_i$, to be the origin;
the center of the outer cylinder, $O_o$, is located at $x=\epsilon_x$, $z=\epsilon_z$.
The inner cylinder of radius $R$ rotates with angular velocity $\omega$; 
the stationary outer cylinder of radius $R+h_0+H_l$ is coated with a soft solid of thickness $H_l \ll R$ 
and Lam\'e coefficients $\lambda$ and $G$.  
Here, $h_0$ is the average distance between the inner 
cylinder and the soft solid.
Following Leal (1992), we use cartesian coordinates to describe the eccentric 
geometry and applied forces, but use polar coordinates to describe the
fluid motion.
When $\frac{h_0}{R} \ll 1$, 
the lubrication approximation reduces the Stokes equations in a cylindrical geometry 
for a fluid of viscosity $\mu$, pressure $p$,
and velocity field ${\bf v} = (v_r, v_\theta)$, to (Leal 1992)
\begin{eqnarray}
\partial_{r} p = 0, \nonumber \\
\label{jb1}
\frac{1}{R}\partial_\theta p = \mu \partial_{rr}v_\theta,
\end{eqnarray}
subject to the boundary conditions 
\begin{eqnarray}
v_r = 0,~~v_\theta=-\omega R~~\hbox{at}~~r=R, \nonumber \\
v_r = 0,~~v_\theta=0~~\hbox{at}~~r=R+h, \nonumber \\
\label{jb4}
p(0) = p(2\pi).
\end{eqnarray}
Since $\frac{h_0}{R} \ll1$, the continuity equation simplifies to (Leal 1992)
\begin{equation}
\label{jb2}
R \partial_r v_r + \partial_\theta v_\theta=0,
\end{equation}
and the gap thickness profile simplifies to
\begin{equation}
\label{jb3}
h(\theta) = h_0 + \epsilon_x \cos \theta + \epsilon_z \sin \theta + H(\theta),
\end{equation}
where $H(\theta)$ is the elastic interface displacement due to the fluid forces.
As in \S \ref{thine},
\begin{equation}
H(\theta) = \frac{H_l\,p(\theta)}{2G + \lambda},
\end{equation}
so that (\ref{jb3}) yields
\begin{eqnarray}
\label{jbndim2}
h = h_0 + \epsilon_x \cos \theta + \epsilon_z \sin \theta +\frac{H_l p}{2G + \lambda}.
\end{eqnarray}
Using the following primed dimensionless variables,
\begin{eqnarray}
v_r = \omega h_0 v_r', ~~~ v_\theta = \omega R v_\theta', ~~~ r = h_0 r', ~~~  h = h_0 h',
 \nonumber \\
\label{jbndim}
p = p^*p' = \frac{\mu R^2\omega}{h_0^2} p',~~~\epsilon_x = h_0\epsilon_x',~~~
\epsilon_z=h_0\epsilon_z',
\end{eqnarray}
we write (\ref{jb1})-(\ref{jb2}), (\ref{jbndim2}), after dropping the primes, as 
\begin{eqnarray}
\partial_{r} p = 0, \nonumber \\
\partial_\theta p = \partial_{rr}v_\theta, \nonumber \\
 \partial_r v_r + \partial_\theta v_\theta=0, \nonumber \\
 \label{preRe}
h  = 1 + \epsilon_x \cos\theta + \epsilon_z\sin\theta + \eta \, p,
\end{eqnarray}
subject to
\begin{eqnarray}
v_r = 0,~~v_\theta=-1~~\hbox{at}~~r=\frac{R}{h_0}, \nonumber \\
v_r = 0,~~v_\theta=0~~\hbox{at}~~r=\frac{R}{h_0}+h, \nonumber \\
p(0) = p(2\pi),
\end{eqnarray}
where,
\begin{equation}
\label{jbeta}
\eta = \frac{H_lp^*}{h_0(2G+\lambda)} = \frac{\mu \omega R^2 H_l}{(2G + \lambda)h_0^3},
\end{equation}
is the softness parameter.
As in \S \ref{ltheory}, we use (\ref{preRe}) to derive the system of equations for the fluid pressure
\begin{eqnarray}
\partial_\theta (6 h + h^3\partial_\theta p)=0, \nonumber \\
h  = 1 + \epsilon_x \cos\theta + \epsilon_z\sin\theta + \eta \, p, \nonumber \\
\label{jbre}
p(0)=p(2\pi).
\end{eqnarray}
In addition, fluid incompressibility implies that  
the average deflection must vanish: 
\begin{equation}
\int_0^{2\pi}h\,d\theta = 2\pi ~~ \to ~~ \int_0^{2\pi} p\,d\theta = 0.
\end{equation}
The forces on the inner cylinder are
\begin{eqnarray}
\label{jbbc}
\int_0^{2\pi} p \sin\theta\, d\theta = L_z, ~~~ \int_0^{2\pi} p \cos\theta\, d\theta = L_x.
\end{eqnarray}
Here, $L_z$ is the vertical force and $L_x$ is the horizontal force.  
We begin with the classical solution for a rigid journal bearing (Leal 1992):
$L_z=1,~L_x=\eta=\epsilon_z=0,~\epsilon_x=0.053$.  
The following brief symmetry argument shows that $\epsilon_z=0$ when $\eta=L_x=0$. 
Since Stokes equations for viscous flow are reversible in time, the transformation $\omega\to-\omega$ 
implies that $\epsilon_z\to-\epsilon_z$.  However, due to symmetry 
we expect the solution to be a reflection about
the $z$-axis and we conclude that $\epsilon_z=0$. 

\begin{figure}
\begin{center}
\includegraphics[width=7cm]{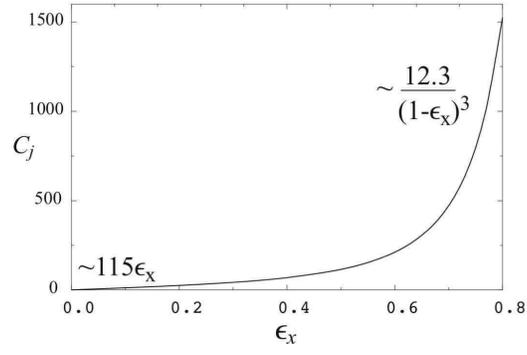}
\par
\caption{For small $\eta$ the dimensionless horizontal force, $L = C_j(\epsilon_x) \eta$, where the
coefficient $C_j$ is plotted above.}
\label{jbLC}
\end{center}
\end{figure}

\begin{figure}
\begin{center}
\includegraphics[width=8cm]{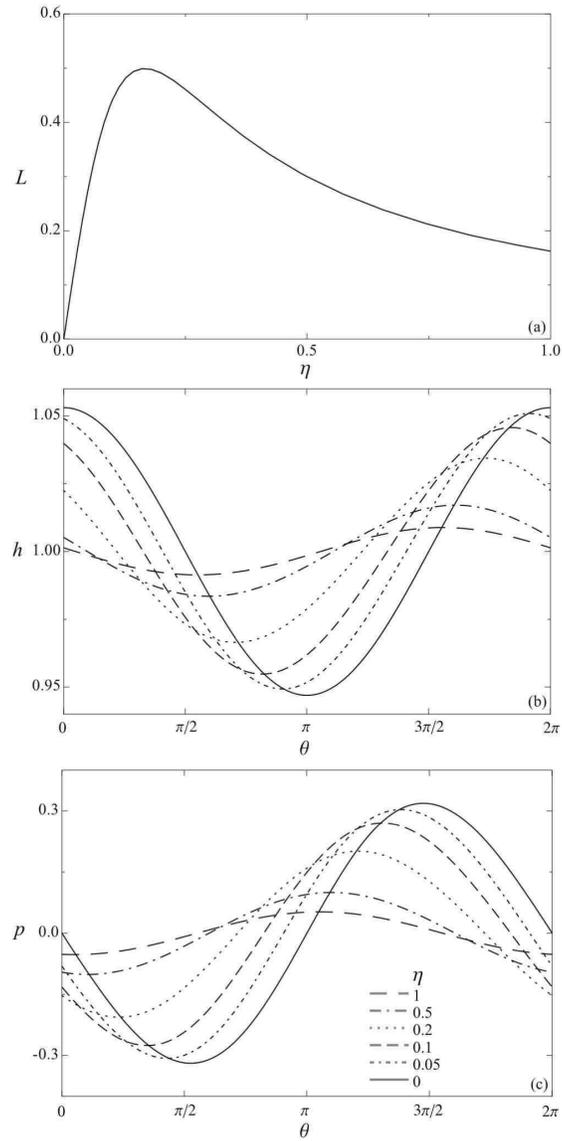}
\par
\caption{(a) Dimensionless horizontal force, $L$, acting on the inner cylinder as a function of $\eta$, 
a measure of the surface deflection; (b) the corresponding gap thickness profiles; 
(c) the corresponding pressure profiles. $\epsilon_z=0$, $\epsilon_x = 0.053$.}
\label{jblift}
\end{center}
\end{figure}

As in previous sections we investigate how elastohydrodynamics alters 
this picture by specifying the eccentricity $(\epsilon_x, \epsilon_z)$ and calculating the forces
$(L_x,L_z)$ as a function of the softness parameter $\eta$.
Solutions to (\ref{jbre})-(\ref{jbbc})
are computed numerically using the continuation software AUTO 2000 
(Doedel {\it et al.} 2004) with $\eta$ as the continuation parameter
and the solution for $\eta=0$ as the initial guess.  
Just as for different geometries analyzed in previous sections,
the deflection of the surface of the soft solid breaks the symmetry and leads to the generation of
a horizontal force in the $x$-direction: $L_x>0$.  
For small deformations ($\eta\ll1$) the dimensionless horizontal force $L=C_j(\epsilon_x) \eta$, where the 
coefficient $C_j(\epsilon_x)$ is shown in figure \ref{jbLC}.  In dimensional 
terms, the horizontal force per unit length for small deformations is 
\begin{equation}
L_x = C_j(\epsilon_x) \frac{p^*H_l}{(2G+\lambda)h_0}p^*R =  C_j(\epsilon_x) \frac{\mu^2 \omega^2 R^5 H_l}{(2G+\lambda)h_0^5}.
\end{equation}
For nearly concentric cylinders, $\epsilon_x \ll 1$, $C_j(\epsilon_x) = 115 \epsilon_x+O(\epsilon_x^2)$.
For large eccentricities, $\epsilon_x \to 1$, the lubrication pressure diverges and 
$C_j(\epsilon_x) \approx 12.3(1-\epsilon_x)^{-3}$.   For $\eta \not\ll 1$ we show $L$, $h$ and $p$
in figure \ref{jblift}.

\section{3-dimensional lubrication flow}
\label{10}

The analysis of the 3-dimensional problem of a sphere moving close to a soft substrate 
is considerably more involved.  
Stone {\it et al.} ({\it in preparation}) are currently engaged in using perturbation
methods to calculate the elastohydrodynamic lift for the case of a sphere translating 
above a thin elastic layer.     
Here, we restrict ourselves to the use scaling arguments to
generalize the quantitative results of previous sections to spherical sliders.
The results are tabulated in Table 2.
In the fluid layer separating the solids, balancing the pressure gradient 
with the viscous stresses yields 
\begin{equation}
\label{pscale}
\frac{p}{l_c} \sim \frac{\mu V}{h^2} \to p \sim \frac{\mu V l_c}{h^2},
\end{equation}
where $l_c$ is the size of the contact zone. 
Substituting $h = h_0 + H_0$ with $H_0 \ll h_0$ we find that the lubrication pressure is 
\begin{equation}
p  \sim  \frac{\mu V l_c}{(h_0+H_0)^2} \sim  \frac{\mu V l_c}{h_0^2}
(1+\frac{h_0}{R}) =   \frac{\mu V l_c}{h_0^2}(1+\eta).
\end{equation}
The reversibility of Stokes equations and the symmetry of paraboloidal contacts 
implies that the lift force $L=0$ when $\eta= \frac{H_0}{h_0}=0$.
For $\eta \ll 1$, we expand the pressure as $p = p^{(0)} + \eta p^{(1)}$.
Since $p = p^{(0)} $ will not generate vertical forces, the 
the lift on a spherical slider, $L_{s}$, 
will scale as
\begin{eqnarray} 
\label{slift}
L_{s} = \eta \int p^{(1)}\, dA \sim \eta \int  \frac{\mu V l_c}{h_0^2}dA \sim  
\eta \frac{\mu V l_c^3}{h_0^2}.
\end{eqnarray}
To compute $L$, we need a prescription for the softness $\eta$ and the contact radius $l_c$ for each 
configuration.

\noindent
(A) For a thin compressible elastic layer (\S \ref{thine}), we substitute $l_c\sim\sqrt{h_0R}$ and 
 $\eta \sim  \frac{\mu (V-\omega R)}{2G + \lambda} \frac{ H_l R^{1/2}}{h_0^{5/2}}$ 
 into (\ref{slift}) to find
\begin{equation}
L_s \sim \frac{\mu^2 V^2H_lR^{2}}{h_0^{3}(2G+\lambda)}.
\end{equation}
(B) For a thin elastic layer with a degenerate axisymmetric conforming contact (\S \ref{Dsec}), 
$l_c \sim (h_0R^{2n-1})^{1/2n}$ 
and $\eta \sim \frac{\mu V}{2G+\lambda} \frac{H_lR^{1-\frac{1}{2n}}}{h_0^{3-\frac{1}{2n}}}$
so that (\ref{slift}) yields 
\begin{equation}
L_{s} \sim \frac{\mu^2 V^2 H_lR^{4-\frac{2}{n}}h_0^{-5+\frac{2}{n}}}{2G + \lambda}
\end{equation}
(C) For a soft spherical slider 
(or thick layer $H_l\gg l_c$; \S \ref{6}), the deflection is given by (\ref{gfcn}) so that 
\begin{eqnarray}
H(x,y) = \frac{\lambda + 2G}{4\pi G(\lambda+G)} \int \frac{p(x',y')dx'\,dy'}{\sqrt{(x'-x)^2+(y'-y)^2}} \sim \frac{p\,l_c}{G}. 
\end{eqnarray}
so that 
\begin{equation}
\label{dh}
\eta \sim  \frac{p\,l_c}{G\, h_0}.
\end{equation}
the size of the contact zone $l_c\sim \sqrt{h_0R}$ so that (\ref{slift}) and (\ref{dh}) yield
\begin{equation}
L_s \sim \frac{\mu^2V^2}{G} \frac{R^{5/2}}{h_0^{5/2}}.
\end{equation}
(D) For an incompressible layer (\S \ref{4}) we have two cases
depending on the thickness of the substrate relative to the 
contact zone characterized by the parameter $\zeta  = \frac{H_l}{l_c}$.
For $\zeta \gtrsim 1$, $l_c \sim \sqrt{h_0R}$ and
$\eta \sim \frac{\mu (V-\omega R)}{G}\frac{R}{h_0^2}$
 so that (\ref{slift}) yields
\begin{equation}
L_{\zeta \gtrsim 1}\sim \frac{\mu^2V^2}{G} \frac{R^{5/2}}{h_0^{5/2}}.
\end{equation}
For the case $\zeta \ll 1$ the proximity of the undeformed substrate substantially
stiffens the layer.  
In sharp contrast to a compressible layer, a thin incompressible layer will deform 
via shear with an effective shear strain
$\frac{\Delta u}{H_l}$.
An incompressible solid must satisfy the
continuity equation $\nabla\cdot{\bf u}=0$, which implies that $\frac{\Delta u}{l_c} \sim \frac{H_0}{H_l}$.
Consequently, 
$\frac{\Delta u}{H_l} \sim \frac{l_cH_0}{H_l^2} $.
Balancing the elastic energy 
$\int G (\frac{l_c H_0}{H_l^2})^2dV 
\sim G (\frac{l_c H_0}{H_l^2})^2 H_l l_c^2$ 
with the work done by the pressure $p H_0 l_c^2$ 
yields 
\begin{equation}
\label{iceta}
\eta \sim \frac{p}{G}\frac{H_l^3}{h_0l_c^2}.
\end{equation}
 Since $l_c \sim \sqrt{h_0R}$, (\ref{pscale}), (\ref{slift}) and 
 (\ref{iceta}) yield
\begin{equation}
\label{18}
\eta \sim \frac{ \mu V}{G} \frac{ H_l^3 }{h_0^{7/2}R^{1/2} }, ~~~ 
L_{\zeta \ll 1} \sim  \frac{\mu^2V^2}{G}\frac{H_l^3 R}{h_0^4}
\end{equation}
(E) For a thin poroelastic layer (\S \ref{5}), $l_c\sim \sqrt{h_0R}$ and 
$\eta \sim \frac{\mu (V-\omega R)}{2G + \lambda} \frac{ H_l R^{1/2}}{h_0^{5/2}}$
so that (\ref{slift}) yields
\begin{equation}
L_s \sim C(\gamma) \frac{\mu^2V^2}{2G+\lambda}\frac{H_lR^{2}}{h_0^{3}},
\end{equation}
where $\gamma$ is the ratio of the poroelastic time scale to the time scale of the motion.
(F) For a spherical shell slider (\S \ref{7}) 
there are two cases: the thickness of the shell, $h_s$,
is smaller than the gap thickness, {\it i.e.} $h_s \ll h_0$ and all the elastic energy is stored in 
stretching; or $h_s  \gtrsim h_0$ and bending and stretching energies are of the same 
order of magnitude (Landau \& Lifshitz 1970).  
For a localized force the deformation 
will be restricted to a region of area $d^2$.
The stretching energy per unit 
area scales as $G h_s H_0^2/R^2$, while the bending energy scales as 
$G h_s^3 H_0^2/d^4$.  The total elastic energy, $U$,  of the deformation is then given by
\begin{equation}
U \approx \frac{G h_s H_0^2 d^2}{R^2}+\frac{G h_s^3 H_0^2}{d^2},
\end{equation}
which has a minimum at $d = \sqrt{h_sR}$.  Comparing $d$ with $l_c \sim \sqrt{h_0R}$ we see that the
hydrodynamic pressure is localized if $h_s  >  h_0$.  For a localized force
$d=\sqrt{h_sR}$ while for a non-localized force $d=R$. The elastic energy of a localized deformation, $U_l$, and a non-localized deformation, $U_n$, are given by
\begin{equation}
\label{ele}
U_l = \frac{G h_s^2 H_0^2}{R}, ~~~ U_n = G h_s H_0^2
\end{equation}
The moment exerted by the hydrodynamic pressure on the spherical shell slider is
\begin{equation}
\label{3Dmom}
M \sim p\,l_c^3,
\end{equation}
which is independent of $h_0$.
The work done by the moment (\ref{3Dmom}), 
which acts through an angle $\Delta\theta \sim H_0/d$, is
\begin{equation}
\label{fwork}
M\Delta\theta \sim p\,l_c^3\frac{H_0}{d}.
\end{equation}
Balancing the work done by the fluid (\ref{fwork}) with the stored 
elastic energy (\ref{ele}) for both nonlocal and local deformations yields
\begin{equation}
\label{hs1}
\eta_n \sim \frac{\mu V\,R}{Gh_sh_0}, ~~~ \eta_l \sim \frac{\mu V\, R^{5/2}}{Gh_s^{5/2}h_0},
\end{equation}
so that (\ref{slift}) and (\ref{hs1}) yield the lift force on the sphere for the two cases
\begin{eqnarray}
L_l \sim \frac{\mu^2 V^2}{G} \frac{R^4}{h_s^{5/2}h_0^{3/2}} ~~~ \hbox{ for } \frac{h_s}{h_0}\gg1, \nonumber 
\\
L_n \sim \frac{\mu^2 V^2}{G} \frac{R^{5/2}}{h_sh_0^{3/2}} ~~~ \hbox{ for } \frac{h_s}{h_0} \lesssim 1.
\end{eqnarray}
(G) For the ball and socket configuration, roughly the 3-dimensional analog of the journal bearing
(\S \ref{8}), $l_c \sim R$ and 
$\eta \sim  \frac{\mu \omega R^2 H_l}{(2G + \lambda)h_0^3}$so that 
the horizontal force 
is given by (\ref{slift}): 
\begin{equation}
L_s \sim  \frac{\mu \omega^2 R^2}{2G+\lambda}\frac{H_lR^4}{h_0^5}.
\end{equation}

\section{Discussion}


The various combinations of geometry and material properties in this paper yield some simple results of great generality : the elastohydrodynamic interaction between soft surfaces immersed in a viscous fluid   leads generically to a coupling between tangential and normal forces regardless of specific material properties or geometrical configurations, {\it i.e.} a lift force  that arises due to the asymmetric fluid pressure deforming the soft solid which  breaks the symmetry of the gap profile. For small surface deformations, $\eta = \frac{\hbox{surface displacement}}{\hbox{characteristic gap thickness}} \ll 1$, the dimensionless normal force is linear in $\eta$.  Increasing $\eta$ ({\it i.e.} softening the material) increases the asymmetry but decreases the magnitude of the pressure.  The competition between symmetry breaking, which dominates for small $\eta$, and decreasing pressure, which dominates for large $\eta$, produces a maximum in the lift force as a function of $\eta$, the material's softness.  


Additional complications such as  nonlinearities and anisotropy in both the fluid and solid, streaming potentials and current generated stresses  (Frank \& Grodzinsky 1987a,b) would clearly change some of our conclusions. However,  the robust nature of the coupling between the tangential and normal forces 
illustrated in this paper should persist and suggests both experiments and design principles for soft lubrication.

\section*{Acknowledgments}
The authors thank Mederic Argentina for assistance using the AUTO 2000 software package,  
and both Howard Stone and Tim Pedley for their thoughtful comments.
We acknowledge support via the Norwegian Research Council (JS), the US Office of Naval Research Young Investigator Program (LM) and the US National Institutes of Health (LM).

\clearpage
\pagebreak

\section*{Figures}

\clearpage

\clearpage

\clearpage

\clearpage

\clearpage

\clearpage

\clearpage

\clearpage

\clearpage

\clearpage

\clearpage

\clearpage

\clearpage

\clearpage

\clearpage

\clearpage

\clearpage

\clearpage

\clearpage

\clearpage

\clearpage



\section*{Tables}

\clearpage

\begin{table}
\caption{Summary of results for small surface deflections. $^*$Upper row corresponds to
$n=2$, while the lower row corresponds to $n=3$ and the undeformed 
dimensionless gap thickness profile is $h=1+x^{2n}$.}
\label{t.1}
\begin{center}
\begin{tabular}{|l|l|c|c|}
\hline
Geometry  & Material & Surface displacement  & Lift force/unit length \\
\hline  \hline
Thin layer  & Compressible  & $\sqrt{2} \frac{\mu V}{2G + \lambda} \frac{ H_l R^{1/2}}{h_0^{3/2}}$ & $\frac{3\sqrt{2}\pi}{4}\frac{\mu^2V^2}{2G+\lambda}\frac{H_lR^{3/2}}{h_0^{7/2}}$ \\
& elastic solid & & 
\\
\hline
Thin layer with & Compressible  & $\frac{\mu V}{2G + \lambda}\frac{ H_lR^{3/4}}{h_0^{7/4}}$
  & $
\frac{351\pi}{784\sqrt{2}}\frac{\mu^2V^2}{2G+\lambda}\frac{H_lR^{9/4}}{h_0^{17/4}}$\\
degenerate contact$^*$ & elastic solid &
$\frac{\mu V}{2G + \lambda}\frac{ H_lR^{5/6}}{h_0^{11/6}}$ & 
$0.8859\frac{\mu^2V^2}{2G+\lambda} \frac{H_lR^{5/2}}{h_0^{9/2}}$
 \\ 
\hline
Soft slider & Elastic solid & 
$ \frac{1}{2\pi}\frac{\mu V(\lambda + 2G)}{G(\lambda+G)}\frac{R}{h_0}$&
$\frac{3\pi^2}{8} \frac{\mu^2 V^2(\lambda + 2G)}{ G(\lambda+G)}\frac{R^2}{h_0^3}$
 \\
\hline
Thickness $\sim\sqrt{Rh_0}$ & Incompressible  & $\frac{1}{2\pi} \frac{\mu V}{G}  \frac{ R}{h_0}$& 
$\frac{C_i(\zeta)}{2\pi} \frac{\mu^2V^2}{G} \frac{R^2}{h_0^3}$
\\
& elastic solid & &\\ 
\hline
Thin layer & Poroelastic &$\sqrt{2}(1-\alpha) \frac{\mu V}{2G + \lambda} \frac{ H_l R^{1/2}}{h_0^{3/2}}$&$
C_p(\gamma)(1-\alpha)\frac{\mu^2V^2}{2G+\lambda}\frac{H_lR^{3/2}}{h_0^{7/2}}$ \\
\hline
Cylindrical shell & Elastic solid & 
$3\sqrt{2}\pi^2 \frac{\mu V(\lambda+2G)}{G(\lambda+G)} \frac{R^{7/2}}{h_s^3h_0^{1/2}} $ &
$6\sqrt{2}\pi^2\,C_s(\frac{h_0}{R}) \frac{\mu^2 V^2(\lambda+2G)}{G(\lambda+G)} \frac{R^{9/2}}{h_s^3h_0^{5/2}}$
 \\
\hline
Journal bearing & Elastic solid &
$ \frac{H_l \mu R^2 \omega}{h_0^2(2G + \lambda)}$
& $C_j(\epsilon_x) \frac{\mu \omega^2 R^2}{2G+\lambda}\frac{H_lR^3}{h_0^5}$ \\
thin layer & & & \\
\hline
\end{tabular}
\end{center}
\end{table}

\begin{table}
\caption{Summary of results for small surface deflections and spherical sliders.}
\label{t.2}
\begin{center}
\begin{tabular}{|l|l|c|}
\hline
Geometry  & Material & Lift force \\
\hline  \hline
Thin layer  & Compressible  & $\frac{\mu^2 V^2H_lR^{2}}{h_0^{3}(2G+\lambda)}$ \\
& elastic solid &  
\\
\hline
Thin layer with & Compressible  & 
$ \frac{\mu^2 V^2 H_lR^{4-\frac{2}{n}}h_0^{-5+\frac{2}{n}}}{2G + \lambda}$\\
degenerate contact & elastic solid & 
\\ 
\hline
Soft slider & Elastic solid &
$ \frac{\mu^2V^2}{G} \frac{R^{5/2}}{h_0^{5/2}}$
 \\
\hline
Thickness $\gtrsim\sqrt{Rh_0}$ & Incompressible  & 
$ \frac{\mu^2 V^2}{G} \frac{R^{5/2}}{h_0^{5/2}}$
\\
& elastic solid  &\\ 
\hline
Thickness $\ll \sqrt{Rh_0}$ & Incompressible  & 
$ \frac{\mu^2V^2}{G} \frac{H_l R^{3/2}}{h_0^{5/2}}$
\\
& elastic solid  &
\\ 
\hline
Thin layer & Poroelastic &
$\frac{\mu^2V^2}{2G+\lambda}\frac{H_lR^{2}}{h_0^{3}}$ \\
\hline
Cylindrical Shell & Elastic solid & 
$\frac{\mu^2 V^2}{G} \frac{R^4}{h_s^{5/2}h_0^{3/2}}$ \\
$h_s\gg h_0$ & &
 \\
\hline
Cylindrical Shell & Elastic solid & 
$ \frac{\mu^2 V^2}{G} \frac{R^{5/2}}{h_sh_0^{3/2}} $ \\
$h_s\lesssim h_0$ & &
 \\
\hline
Journal bearing & elastic solid &  $\frac{\mu \omega^2 R^2}{2G+\lambda}\frac{H_lR^4}{h_0^5}$ \\
thin layer & & \\
\hline
\end{tabular}
\end{center}
\end{table}

 \end{document}